\documentclass[12pt]{article}
\usepackage{graphicx}
\usepackage{epsfig}
\begin{document}
\newcommand{\beq}{\begin{equation}}
\newcommand{\eeq}{\end{equation}}
\newcommand{\bea}{\begin{eqnarray}}
\newcommand{\eea}{\end{eqnarray}}
\newcommand{\eps}{\varepsilon}
\newcommand{\Fs}{\mbox{\scriptsize F}}
\newcommand{\inn}{\mbox{\scriptsize in}}
\newcommand{\exx}{\mbox{\scriptsize ex}}
\newcommand{\lsim}{\stackrel{\scriptstyle <}{\phantom{}_{\sim}}}
\newcommand{\gsim}{\stackrel{\scriptstyle >}{\phantom{}_{\sim}}}

{\bf A realistic model of superfluidity in the neutron star inner
crust.}

\vskip 0.5 cm \noindent M.~Baldo$^{1}$, E.E.~Saperstein$^{2}$ and
S.V.~Tolokonnikov$^{2}$

\vskip 0.5 cm

\noindent $^1$INFN, Sezione di Catania, 64 Via S.-Sofia, I-95123
Catania,
Italy \\
$^{2}$ Kurchatov Institute, 123182, Moscow, Russia

\vskip 0.5 cm

\begin{abstract}

A semi-microscopic self-consistent quantum  approach developed
recently to describe the inner crust structure of neutron stars
within the  Wigner-Seitz (WS) method with the explicit inclusion of
neutron and proton pairing correlations is further developed. In
this approach, the generalized energy functional is used which
contains the anomalous term describing the pairing. It is
constructed by matching the realistic phenomenological functional by
Fayans et al. for describing the nuclear-type cluster in the center
of the WS cell with the one calculated microscopically for neutron
matter. Previously the anomalous part of the latter was calculated
within the BCS approximation. In this work corrections to the BCS
theory which are known from the many-body theory of pairing in
neutron matter are included into the energy functional in an
approximate way. These modifications have a sizable influence on the
equilibrium configuration of the inner crust, i.e.  on the proton
charge $Z$ and the radius $R_c$ of the WS cell. The effects are
quite significant in the region where the neutron pairing gap is
larger.

\end{abstract}

in hadronic systems and effective interactions

\section{Introduction}

Recently, we have developed a semi-microscopic self-consistent
approach to describe the neutron star inner crust with the explicit
inclusion of neutron and proton pairing correlations
\cite{crust3,crust4}. The inner crust is the  part of the neutron
star shell with subnuclear densities, $0.001 \rho_0 \le \rho \le 0.5
\rho_0 $, where $\rho_0$ is the normal nuclear density. It is a
crystal system consisting mainly of spherically symmetrical
nuclear-like clusters immersed in a sea of neutrons and virtually
uniform sea of electrons. The quantum self-consistent description of
the inner crust goes back to the classical paper by Negele and
Vautherin (N\&V)\cite{NV} who used a kind of energy functional
method combined with the Wigner-Seitz (WS) method to describe the
crystal structure effects in an approximate way.
 Within this
approach, for a fixed average nuclear density $\rho$, the energy
functional of the system is minimized for the spherical WS cell
 of the radius $R_c$.
A cell contains $A=(4\pi/3)R_c^3\rho$ nucleons, specifically  $Z$
protons and $N=A-Z$ neutrons. As far as the system is
electro-neutral, the number of electrons per a cell is equal to $Z$.
For a mature neutron star which can be considered at zero
temperature and neutrino free, the $\beta$-stability condition
consists in equality of the neutron chemical potential $\mu_n$ to
the sum of the proton and electron ones, $\mu_p{+} \mu_e$. For a
wide region of $\rho$, the minimization procedure was carried out in
\cite{NV} for different values of $Z$ and $R_c$ resulting in the
equilibrium configuration ($Z,R_c$) for the density under
consideration.

It should be noted that the pairing effects were not taken into
account in \cite{NV}. Thus, the approach developed in
\cite{crust3,crust4} could be considered as a  generalization of
the N\&V method with allowance for pairing effects. Although the
contribution of the pairing to the total binding energy of the
system under consideration is rather small, it turned out that it
may to change the equilibrium configuration ($Z,R_c$)
significantly.

To involve the pairing effects in a self-consistent way, we use
the generalized energy functional method  \cite{Fay} which
incorporates in a natural way the pairing into the original
Kohn-Sham \cite{KS} method. In this approach, the interaction part
of the generalized energy functional is the sum of  the normal
component, ${\cal E}_{\rm norm}(\rho)$, and the anomalous one,
${\cal E}_{\rm an}(\rho, \nu)$. They depend on the normal
densities ($\rho_n , \rho_p$) and the anomalous ones, ($\nu_n,
\nu_p$). Just as in the Kohn-Sham method, the prescription $m^*=m$
holds to be true. The so-called Superfluid LDA method suggested
recently \cite{Bulgac1} is rather close to the method by S. Fayans
et al. The main difference between the two approaches is in the
form of the density dependence of the anomalous term of the energy
functional.

 In \cite{crust3,crust4} the semi-microscopic
energy functional was constructed with matching at the nuclear
cluster surface the phenomenological nuclear  functional ${\cal
E}^{\rm ph}$ by Fayans et al. \cite{Fay} inside the cluster to a
microscopic one, ${\cal E}^{\rm mi}$, for the neutron environment.
The normal part of ${\cal E}^{\rm mi}$ was found in \cite{B-V}
within the Brueckner approach with the Argonne  v$_{18}$ potential
\cite{v18} . The anomalous component of ${\cal E}^{\rm mi}$ was
calculated in \cite{crust3,crust4} within the
Bardeen-Cooper-Schrieffer (BCS) approximation for neutron matter,
again with the use of the v$_{18}$ potential.

It is well known that the BCS approximation overestimates the gap
value $\Delta_n$ in neutron matter. Various many-body corrections
suppress the value of $\Delta_n^{\rm BCS}$  significantly.
Although up to now there is no consistent many-body theory of
pairing in neutron matter, there exists a conventional point of
view
 (see e.g. \cite{crust2}) that the BCS gap value is suppressed,
\beq \Delta_n(k,k_{\Fs}) = f_{\rm m-b}(k,k_{\Fs}) \Delta_n^{\rm
BCS}(k,k_{\Fs}), \label{fac} \eeq by a factor $f_{\rm
m-b}(k,k_{\Fs})$ which is between 1/2 and 1/3. The only exception
seems to be the work in ref. \cite{adel}. In this paper, we use a
simple model for the many-body corrections in which the
suppression factor $f_{\rm m-b}(k,k_{\Fs})$ is supposed to be
momentum and density independent.  This ansatz is essentially
similar to that used in \cite{Bulgac2} for considering the
structure of a superfluid vortex in neutron matter. We modify the
anomalous component of the energy functional of
\cite{crust3,crust4} by introducing the constant suppression
factor. We examine two versions of this model, the P2 model
($f_{\rm m-b}=1/2$) and the P3 one ($f_{\rm m-b}=1/3$). In this
notation, it is natural to name the BCS approximation ($f_{\rm
m-b}=1$) as the P1 model. We expect that the real truth is
somewhere between the P2 and P3 models.

One more remark should be made before going to the body of the
article. As it was found recently \cite{BC12}, there are internal
uncertainties inherent to the WS  method applied to the neutron star
inner crust. They originate from the kind of the boundary conditions
for the single-particle functions used in the WS method. There are
two kinds of such boundary conditions which {\it a priori} seem
equivalent. As it turned out, the corresponding predictions for the
equilibrium configurations ($Z,R_c$) are in general different. As a
rule, the difference is not large, corresponding to variation of $Z$
by 2 -- 6 units and of $R_c$, by 1 -- 2 fm. However, sometimes
strong changes in the neutron single-particle spectrum arise which
influence the solution of the gap equation significantly. Therefore,
for each model under consideration, we carried out calculations for
both kinds of boundary conditions.

\section{ Modification of the BCS anomalous part of the Generalized
Energy Functional}

The ansatz of \cite{crust3,crust4} for the complete energy
functional consists in a smooth matching of the phenomenological
and the microscopic functionals at the cluster surface:
 \beq
{\cal E}(\rho_{\tau}({\bf r}),\nu_{\tau}({\bf r})) = {\cal E}^{\rm
ph}(\rho_{\tau}({\bf r}),\nu_{\tau}({\bf r})) + \left({\cal E}^{\rm
mi}(\rho_{\tau}({\bf r}),\nu_{\tau}({\bf r}))-{\cal E}^{\rm
ph}(\rho_{\tau}({\bf r}),\nu_{\tau}({\bf r}))\right)\left(1 -
F_m(r)\right), \label{tot} \eeq where $\tau=n,p$ is the isotopic
index and the matching function $F_m(r)$ is a two-parameter Fermi
function. The latter is taken to be the same for the normal and the
anomalous components of the energy functional, with the diffuseness
parameter $d_m{=}0.3\;$fm and the matching radius $R_m$ which should
be chosen anew in any new case, in accordance with the equality of $
\rho_p(R_m)=0.1 \rho_p(0)$. For such a choice, practically all the
protons are located inside the radius $R_m$. Therefore, the matching
procedure concerns,in fact, only neutrons, protons being described
with the pure phenomenological nuclear energy functional. In
practice, we use in Eq.~(\ref{tot})  an approximation in which only
neutron components of the microscopic and phenomenological
functionals are taken into account in the second term containing the
difference of (${\cal E}^{\rm mi}-{\cal E}^{\rm ph}$).

Following to \cite{crust3,crust4}, we use for the microscopic part
of the normal component of the total energy functional (\ref{tot})
the one  calculated in \cite{B-V} for neutron matter with the
Argonne v$_{18}$ potential.  Its explicit form could be found in
the cited articles. Here we concentrate on the anomalous part of
the energy functional which will be modified in comparison with
that of \cite{crust3,crust4}.

The anomalous part of the energy functional  used in
\cite{crust3,crust4} has the form: \beq {\cal E}_{\rm an} =\frac
{1}{2} \sum_{\tau} {\cal V}_{{\rm an},\tau}^{\rm eff}(r)
|\nu_{\tau}({\bf r})|^2, \label{an} \eeq where ${\cal V}_{{\rm
an},\tau}^{\rm eff}$ is the density dependent effective pairing
interaction.

 The matching relation (\ref{tot}) for the anomalous part of the
 energy functional leads to the analogous relation for the effective
pairing interaction:
 \beq {\cal V}_{\rm an}^{\rm eff}(r)= {\cal V}_{\rm eff}^{\rm ph}(\rho(r)) F_m( r) +
 {\cal V}_{\rm eff}^{\rm mi}(\rho(r))(1-F_m( r)).
\label{mEPI} \eeq The isotopic index $\tau$ is for brevity
omitted.

We shall use the same phenomenological effective pairing
interaction ${\cal V}_{\rm eff}^{\rm ph}$ as in
\cite{crust3,crust4}, therefore we omit here its explicit form.
Let us note only that it has a density dependent coordinate
delta-function form of \cite{Fay}.  The explicit form of the
density dependence \cite{Fay} was modified a little in
\cite{crust3} in accordance with (\ref{mEPI}). As to the
microscopic effective pairing interaction, it was calculated in
\cite{crust3} microscopically within the BCS approximation with
the same Argonne force v$_{18}$ as the normal part of the energy
functional . In this paper, we  modify this procedure to take into
account approximately the many-body corrections to the BCS
approximation. Let us first repeat the BCS procedure.

The microscopic part of the effective pairing interaction, ${\cal
V}_{\rm eff}^{\rm mi}(r)$, should be found for the model space
$S_0$ under consideration which is limited with the energy $E_0$
for the single-particle spectrum. For a fixed value of the neutron
density $\rho_n(r)$, it is defined via the gap equation in
homogeneous neutron matter with the density $\rho=\rho_n(r)$. We
start from the  BCS approximation in which the gap $\Delta$ is
expressed directly in terms of the bare NN potential $v(k,k')$ in
the $1S_0$ channel: \beq
 \Delta(k) \, =\, - \sum_{k'} v(k,k') \frac {\Delta(k')} {2E(k')},
\label{Delta0} \eeq where $E(k) = \sqrt{(\eps_k - \mu_n)^2
+\Delta^2(k)}$, $\eps_k=k^2/2m+U_n$, $U_n$ is the value of the
neutron matter potential well. In terms of the effective pairing
interaction, the gap equation looks analogously, but the
integration in the momentum space is limited within the model
space $S_0$: \beq
  \Delta(k) \, =\, - \sum_{k'<k_0}{\cal V}_{\rm eff}^{\rm mi} (k,k') \frac
  {\Delta(k')}  { 2E(k')},
\label{Delta} \eeq where $k_0=\sqrt{2m(E_0+\mu_n-U_n)}$.

In the BCS approximation the relation between the effective
pairing interaction and the bare NN potential is obvious: \beq
{\cal V}_{\rm eff}^{\rm mi} (k,k') \, =\, v(k,k') - \sum_{k_1>k_0}
v(k,k_1) \frac {1} {2E(k_1)}
  {\cal V}_{\rm eff}^{\rm mi} (k_1,k').
\label{EPI} \eeq

The effective pairing interaction entering Eq.(\ref{Delta})
depends explicitly on momenta,
 which corresponds to a non-local force in the
coordinate space. In view of very simple local form  of the
phenomenological effective pairing interaction ${\cal V}_{\rm
eff}^{\rm ph}$ in Eq. (\ref{mEPI}), for matching it is necessary
to simplify the microscopic partner ${\cal V}_{\rm eff}^{\rm mi}$
to a local form, too. The simplest way is, for a fixed value of
$\rho$ under consideration, to define it as a $k$-independent
average value of the effective pairing interaction in Eq.
(\ref{Delta}) which yields the same value $\Delta(k_{\Fs})$ as the
exact effective pairing interaction:

\beq
  \Delta(k_{\Fs}) \, =\, - \bar {{\cal V}}_{\rm eff}^{\rm mi}(k_{\Fs}) \sum_{k'<k_0} \frac
  {\Delta(k')}  { 2E_0(k')},
\label{DeltaF} \eeq where $k_{\Fs}$ is the local Fermi momentum
$k_{\Fs}=(3 \pi^2 \rho)^{1/3}$, and $E_0(k) = \sqrt{(\eps_k -
\mu_n)^2 + \Delta^2(k_{\Fs})}$. For the Fermi momentum $k_{\Fs}$
under consideration, the microscopically calculated value of
$\Delta(k_{\Fs})$ is the only input of Eq.~(\ref{DeltaF}) to find
the effective pairing interaction $\bar {{\cal V}}_{\rm eff}^{\rm
mi}(k_{\Fs})$. In the BCS (or P1) model, one uses $\Delta^{\rm
BCS}(k_{\Fs})$ in Eq.~(\ref{DeltaF}). In the P2 or P3 models the
value of $\Delta(k_{\Fs})$ is  found from Eq.~(\ref{fac}) with
$f_{\rm m-b}{=}1/2$ or 1/3, correspondingly. Then it should be
substituted  into Eq.~(\ref{DeltaF}).
 The resulting values of the effective pairing interaction in
 neutron matter for the case of
the P1 (BCS), P2 and P3 models are displayed in Fig.~1.

\begin{figure}
\centerline{\includegraphics [height=100mm,width=120mm]{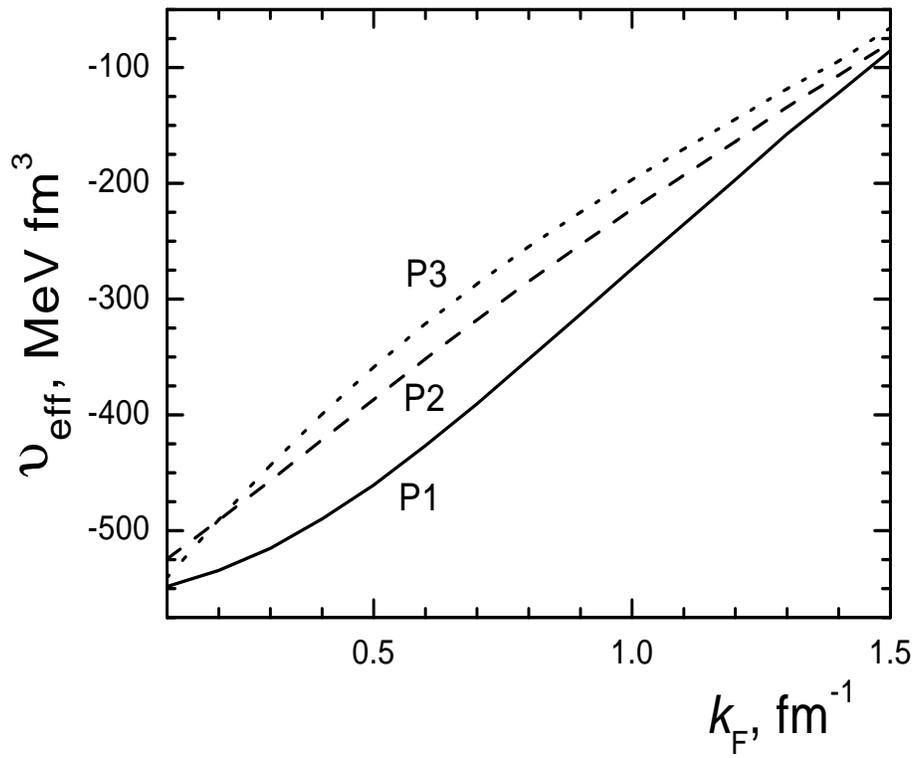}}
\vspace{2mm} \caption{ The effective pairing interaction in
neutron matter  for the P1 , P2  and P3 models (solid, dashed and
dotted line, correspondingly).}
\end{figure}

Let us now discuss the problem of the boundary conditions in the
WS method mentioned above. For the case of the BCS approximation,
it was examined in \cite{BC12}. Application of the variational
principle to the energy functional under consideration for a WS
cell results in the set of the Shr\"odinger-type equations for the
single particle neutron functions $\phi_{\lambda}({\bf
r})=R_{nlj}(r) \Phi_{ljm}({\bf n})$, with the standard notation.
The radial functions $R_{nlj}(r)$ obey the boundary conditions at
the point $r=R_c$. There exist different kinds of the boundary
conditions. N\&V  used the following one:
 \beq
 R_{nlj}(r=R_c)=0
 \label{bco}
\eeq for odd $l$, and \beq   R'_{nlj}(r=R_c)=0, \label{bce} \eeq
 for even ones.
In \cite{BC12} it was denoted  as BC1. An alternative kind of the
boundary conditions (BC2) was considered also there, when
Eq.~(\ref{bco}) is valid for even $l$ whereas Eq.~(\ref{bce}), for
odd ones. As the analysis of \cite{BC12} for the BCS case (P1
model) has shown, some predictions of the two versions of the
boundary conditions (BC1 {\it versus} BC2) are in general
different. In this paper, we will perform the similar analysis for
the P2 and P3 models.

\section{ A brief summary of the pairing effects in the case of the
BCS approximation}

 Calculations of \cite{crust3,crust4} were carried with
the use of the  BCS approximation for the neutron matter pairing and
the N\&V  version of the boundary conditions (BC1, in our notation).
As it turned out, the pairing correlations influence the equilibrium
values of ($Z,R_c$) significantly.  In the paper \cite{BC12}, for
the case of the BCS  approximation (i.e., the P1 model), the
dependence of the ground state properties of the inner crust on the
kind of the WS boundary conditions was examined. The calculations
were carried out directly with the two kinds of boundary conditions,
BC1 and BC2, and the results were compared with each other. To make
the analogous analysis for the P2 and P3 models in the next sections
more transparent, we cite here some results of \cite{BC12}. In
particular, Fig.~2 shows the values of the binding energy per a
nucleon, $E_{\rm B}$, for the P1 model calculated with the two
versions of the boundary conditions.

\begin{figure}
\centerline{\includegraphics [height=180mm,width=100mm]{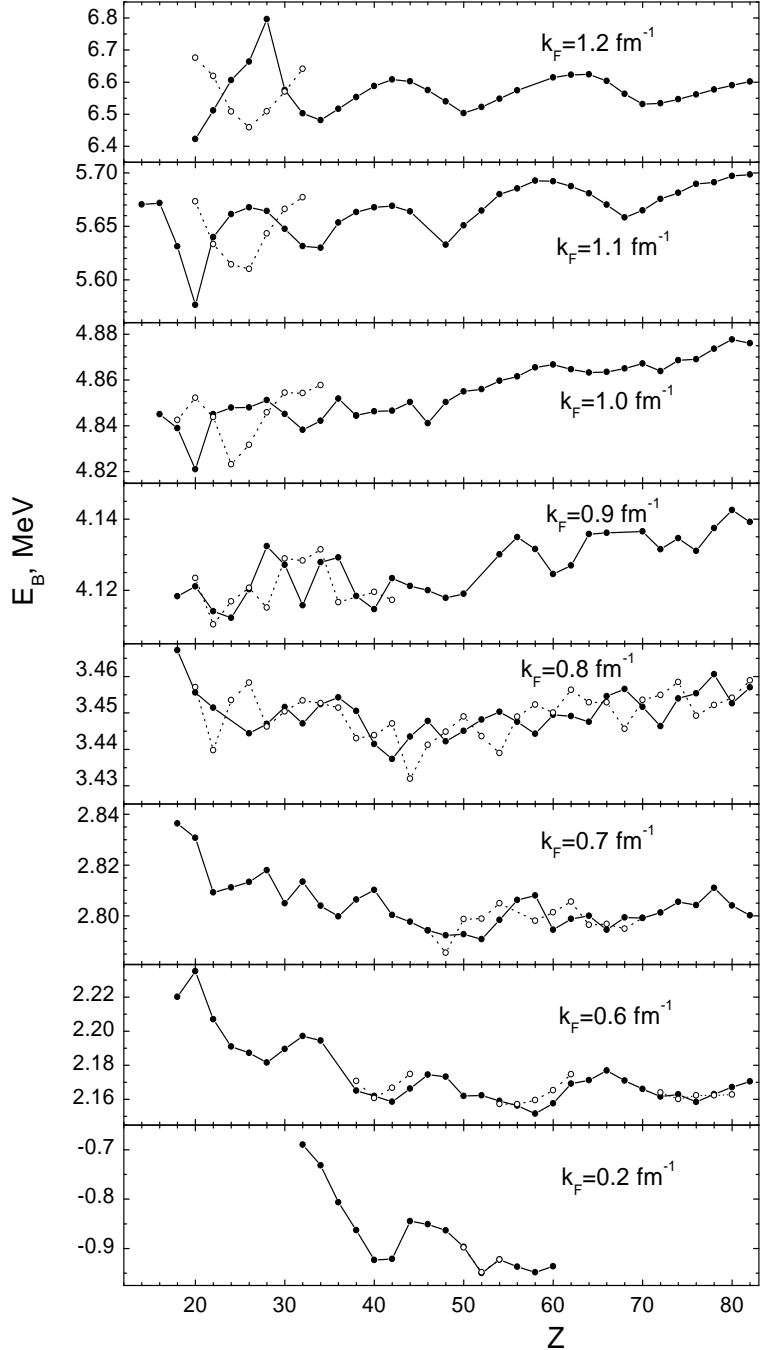}}%
\vspace{0mm} \caption{ Binding energy per a nucleon for various
$k_{\Fs}$ in the BC1 case  (solid circles connected with the solid
lines) and the BC2 one (open  circles connected with the dotted
lines).}
\end{figure}

We see that in the case of very small density, $k_{\Fs}{=}
0.2\;$fm$^{-1}$, which is nearby the neutron drip point, the
predictions of the BC1 and BC2 versions are practically
 identical.  At increasing density, with $k_{\Fs} \ge 0.6\;$fm$^{-1}$,
 the  uncertainty in the equilibrium value of $Z$ is between 2 and
 6 units, with the largest values at the largest $k_{\Fs}$. The
 uncertainty in the value of $R_c$ is, as a rule, about 1 fm and
 only for $k_{\Fs}{=}1.1\;$fm$^{-1}$ it turns out to be about 2
fm. Referring  to  \cite{crust4,BC12} for more details, we present
in Table 1 the main ground state characteristics of the inner
crust. There are two lines for each value of $k_{\Fs}$. The first
one is given for the $Z$ value corresponding to the minimum of the
binding energy $E_{\rm B}$ in the BC1 case, the second one, for
BC2. The only exception is $k_{\Fs}{=}0.2\;$fm$^{-1}$ where these
two values of $Z$ coincide. It should be stressed that, as a rule,
the value of these uncertainties is  smaller than the variations
of the equilibrium configuration ($Z,R_c$) connected with  the
pairing effects \cite{crust4}. One more point which is important
for the analogous consideration within the P2 and P3 models is as
follows. For all the values of $k_{\Fs}$ which were investigated,
the relative position of the local minima  of the functions
$E_{\rm B}(Z)$ is always similar for the BC1 case and the BC2 one.
As the result, the corresponding  absolute minima are rather close
to each other.

\begin{table}
\caption{Comparison of characteristics of equilibrium
configurations of the WS cell for two different kinds of the
boundary conditions in the case of the P1 model (BCS
approximation).}
\bigskip
\begin{center}
\begin{tabular}{|c|c|c|c|c|c|c|c|}
\hline
  {$k_{\rm F},$} & \raisebox{-6pt}{$\,Z\,$} &
  \multicolumn{2}{|c|}{$R_{\rm c},\;$fm}\rule{0pt}{14pt}&
  \multicolumn{2}{|c|}{$E_{\rm B},\;$MeV}&
  \multicolumn{2}{|c|}{$\mu_n,\;$MeV}
  \\
\cline{3-8}
 \rule{0pt}{13pt} ${\rm fm^{-1}}$&& BC1 & BC2 & BC1 & BC2 & BC1 & BC2\\
\hline\rule{0pt}{14pt}
 0.2  & 52 & 57.18 & 57.10 &-0.9501 &-0.9483 & 0.1928 & 0.1942\\
\hline
  \raisebox{-6pt}{0.6} & 58 & 37.51 & 37.48 & 2.1516 & 2.1596 & 3.2074 & 3.2226 \\
      & 56 & 36.97 & 36.95 & 2.1563 & 2.1572 & 3.2173 & 3.2193    \\
\hline
  \raisebox{-6pt}{0.7} & 52 & 32.02 & 32.04 & 2.7908 & 2.7989 & 3.9876 & 4.0107 \\
      & 48 & 31.16 & 31.14 & 2.7924 & 2.7856 & 4.0069 & 3.9873   \\
\hline
  \raisebox{-6pt}{0.8} & 42 & 26.90 & 26.91 & 3.4373 & 3.4471 & 4.8454 & 4.8561 \\
      & 44 & 27.29 & 27.30 & 3.4435 & 3.4319 & 4.8553 & 4.8198   \\
\hline
  \raisebox{-6pt}{0.9} & 24 & 20.26 & 20.30 & 4.1123 & 4.1169 & 5.7340 & 5.7986 \\
     &  22 & 19.87 & 19.70 & 4.1141 & 4.1104 & 5.7861 & 5.7170    \\
\hline
  \raisebox{-6pt}{1.0} & 20 & 16.69 & 16.90 & 4.8210 & 4.8522 & 6.8525 & 6.7424 \\
      & 24 & 18.29 & 18.22 & 4.8479 & 4.8231 & 6.8446 & 6.8920    \\
\hline
  \raisebox{-6pt}{1.1} & 20 & 14.99 & 15.33 & 5.5765 & 5.6733 & 7.4288 & 8.0446 \\
      & 26 & 16.75 & 17.08 & 5.6677 & 5.6100 & 7.9680 & 8.5398     \\
\hline
  \raisebox{-6pt}{1.2} & 20 & 13.68 & 13.95 & 6.4225 & 6.6762 & 8.5814 & 9.1898 \\
      & 26 & 15.21 & 14.89 & 6.6639 & 6.4587 & 9.0825 & 9.3413   \\[2pt]
\hline
\end{tabular}
\end{center}
\end{table}

Table 2 collects some important characteristics of the gap for all
the density values under consideration. The relative position of
two lines for the same value of $k_{\Fs}$ is the same as in Table
1, i.e. the first line corresponds to Z from the equilibrium
($Z,R_c$) configuration for the BC1 version. The following
notation is used. The asymptotic value of the Fermi momentum
$k_{\rm F}^{\rm as}$ corresponds to the asymptotic value of the
density $\rho(r)$ averaged over the interval $R_c{-}b < r < R_c $,
$b{=}2\;$fm. The asymptotic gap value $\Delta_{\rm as}$ is found
as the average  of $\Delta(r)$ over the same interval. The central
gap value $\Delta(0)$ is calculated as the average of $\Delta(r)$
over the interval $0<r<3\;$fm. The Fermi average gap
$\Delta_{\Fs}$ is defined as the average value  of the diagonal
matrix element of the neutron gap at the Fermi surface. The
averaging procedure involves 10 levels above $\mu_n$ and 10 levels
below.  At last, $\Delta_{\rm inf}$ means the infinite neutron
matter gap value found within the BCS approximation for the
density $\rho$ corresponding to the Fermi momentum $k_{\rm F}^{\rm
as}$, and $\Delta^0_{\rm inf}$ is the same for the Fermi momentum
$k_{\rm F}$. Let us remind that the latter corresponds to the
average nucleon density under consideration. Obviously, the
inequality $k_{\rm F}^{\rm as} < k_{\rm F}$ takes place because
the WS cell contains a nuclear-like cluster in the center with the
density which exceeds the average one. The difference is
especially large in the case of $k_{\rm F}{=}0.2\;$fm which is
nearby the neutron drip point. Indeed, in this case almost all the
matter is concentrated in the central blob. So big difference
between the values of $\Delta_{\rm inf}$ and $\Delta^0_{\rm inf}$
for $k_{\rm F}{=}0.2\;$fm is explained, first, by the big
difference between two values of the Fermi momentum and, second,
with the sharp dependence of the $\Delta_n$ in neutron matter on
$k_{\Fs}$ at small $k_{\Fs}$.

It is worth to mention that the difference between the asymptotic
$\Delta_{\rm as}$ value and the infinite neutron matter prediction
$\Delta_{\rm inf}$ is a measure of validity of the LDA for the gap
calculation outside the central nuclear cluster. One can see that,
as a rule, the LDA works within 10\% accuracy, but sometimes the
difference is greater which is an evidence of the so-called
proximity effect. The Fermi average value $\Delta_{\Fs}$ is
usually very close to $\Delta_{\rm as}$ value. It is explained
with the fact that the region out of the nuclear cluster, in which
the function $\Delta(r)$ is almost a constant, contributes mainly
to the matrix elements of $\Delta$ nearby the Fermi surface.

\begin{table}
\caption{Average gap characteristics in the P1 model (BCS
approximation). }
\bigskip
\begin{tabular}{|c|c|c|c|c|c|c|c|c|c|c|c|c|}
\hline\rule{0pt}{14pt}
  {$k_{\rm F},\!\!$} & \raisebox{-6pt}{$Z$} &
  \multicolumn{2}{|c|}{$k_{\rm F}^{\rm as},\,{\rm fm^{-1}}$}&
  \multicolumn{2}{|c|}{$\Delta(0),\;$MeV}&
  \multicolumn{2}{|c|}{$\Delta_{\rm as},\;$MeV}&
  \multicolumn{2}{|c|}{$\Delta_{\rm F},\;$MeV}&
  \multicolumn{2}{|c|}{$\Delta_{\rm inf},\;$MeV}&
  {$\Delta^0_{\rm inf},\!\!$}\\
\cline{3-12}
 \rule{0pt}{13pt}$\rm fm^{-1}\!\!$&& BC1 & BC2 & BC1 & BC2 & BC1 & BC2 & BC1 & BC2 & BC1 &  BC2
 &MeV$\!\!$\\
\hline\rule{0pt}{14pt} 0.2 & 52 & 0.1156 & 0.1095  &0.088 &0.132
&0.042 &0.046 &0.043 &0.058 & 0.126 &0.106 &0.40\\
 \hline \raisebox{-6pt}{0.6} & 58 &0.5786 &0.5783 &1.464 & 1.471 & 1.947 & 1.899 &1.919
&1.893 & 2.321 &2.320 &  \raisebox{-6pt}{2.42}\\
& 56 & 0.5783 &0.5786 &1.456 &1.428 &1.899 &1.912
&1.893 &1.891 &2.319 &2.321 &\\

\hline \raisebox{-6pt}{0.7}& 52 & 0.6758 & 0.6753 &1.665 &1.650
&2.358 &2.288 &2.300 &2.247 &2.680 &2.678 &  \raisebox{-6pt}{2.76}\\
 & 48 &0.6763 &0.6763 &1.679 &1.648 &2.312 &2.368
&2.290 &2.325 &2.682 &2.682 & \\

\hline \raisebox{-6pt}{0.8} &42 &0.7732 & 0.7724 & 1.767 &1.726
&2.614 &2.546
&2.555 &2.445  &2.883 &2.882 &\raisebox{-6pt}{2.93}\\
& 44 &0.7729 &0.7727 &1.747 &1.834 &2.580 &2.679 &2.525 &2.560 &2.883 &2.883 & \\
\hline \raisebox{-6pt}{0.9}& 24 &0.8694 &0.8693 &1.862 &1.664
&2.777 &2.625 &2.636 &2.506 &2.919 &2.919 & \raisebox{-6pt}{2.92}\\
 &22 & 0.8725 &0.8664 &1.936 &1.654 &2.677 &2.680
&2.617 &2.544 &2.918 &2.919 &\\

\hline \raisebox{-6pt}{1.0} &20 &0.9499 &0.9613 &1.249 &1.966
&2.199 &2.635
&2.023 &2.517 &2.800 &2.773 & \raisebox{-6pt}{2.68} \\
& 24 &0.9612` &0.9574 &1.894 &1.504 &2.705 &2.507 &2.519 &2.288 &2.774 &2.782 &\\
\hline \raisebox{-6pt}{1.1} &20 &1.0315 &1.0531 &0.996 &1.889
&1.477 &2.411
&1.318 & 2.317 &2.550 &2.458 &\raisebox{-6pt}{2.26}\\
& 26 &1.0434 &1.0649 &1.927 &1.296 &2.469 &2.242 &2.280 &2.020 &2.500 &2.408 & \\
\hline \raisebox{-6pt}{1.2} &20 &1.1243 &1.1321 &1.556 &0.992
&1.340 &2.017
&1.210 &1.558 &2.113 &2.066 &\raisebox{-6pt}{1.66}\\
& 26 &1.1278 &1.1160 &0.760 &0.991 &1.549 & 0.963 &1.249 &0.862
&2.092 &2.163&\\[2pt]
 \hline
\end{tabular}
\end{table}

For the case of small and intermediate densities, $k_{\Fs} <
1\;$fm$^{-1}$, the influence  of a particular choice of the
boundary conditions, BC1 or BC2, to the value of $\Delta_{\Fs}$ or
$\Delta_{\rm as}$ is not essential. As the result, the uncertainty
in predictions for the gap function $\Delta(r)$ caused by this
choice of the boundary conditions is also rather small. An example
 for $k_{\rm F}{=}0.8\;$fm is given in Fig. 3. The difference
between any couple of these curves is less than the accuracy of
the approach, and any of them could be used as a prediction for
$\Delta(r)$.  To be definite, let us consider the
``self-consistent''  gap function for the BC1 version of the
boundary conditions  as the prediction of the WS method for
$\Delta(r)$ in the case of small and intermediate densities,
$k_{\Fs}< 1.0\;$fm$^{-1}$. In the case of
$k_{\Fs}{=}0.8\;$fm$^{-1}$ under consideration it corresponds to
$Z{=}42$. Such a choice is similar to that used in \cite{crust4}.

\begin{figure}[!h]
\centerline{\includegraphics [height=100mm,width=120mm]{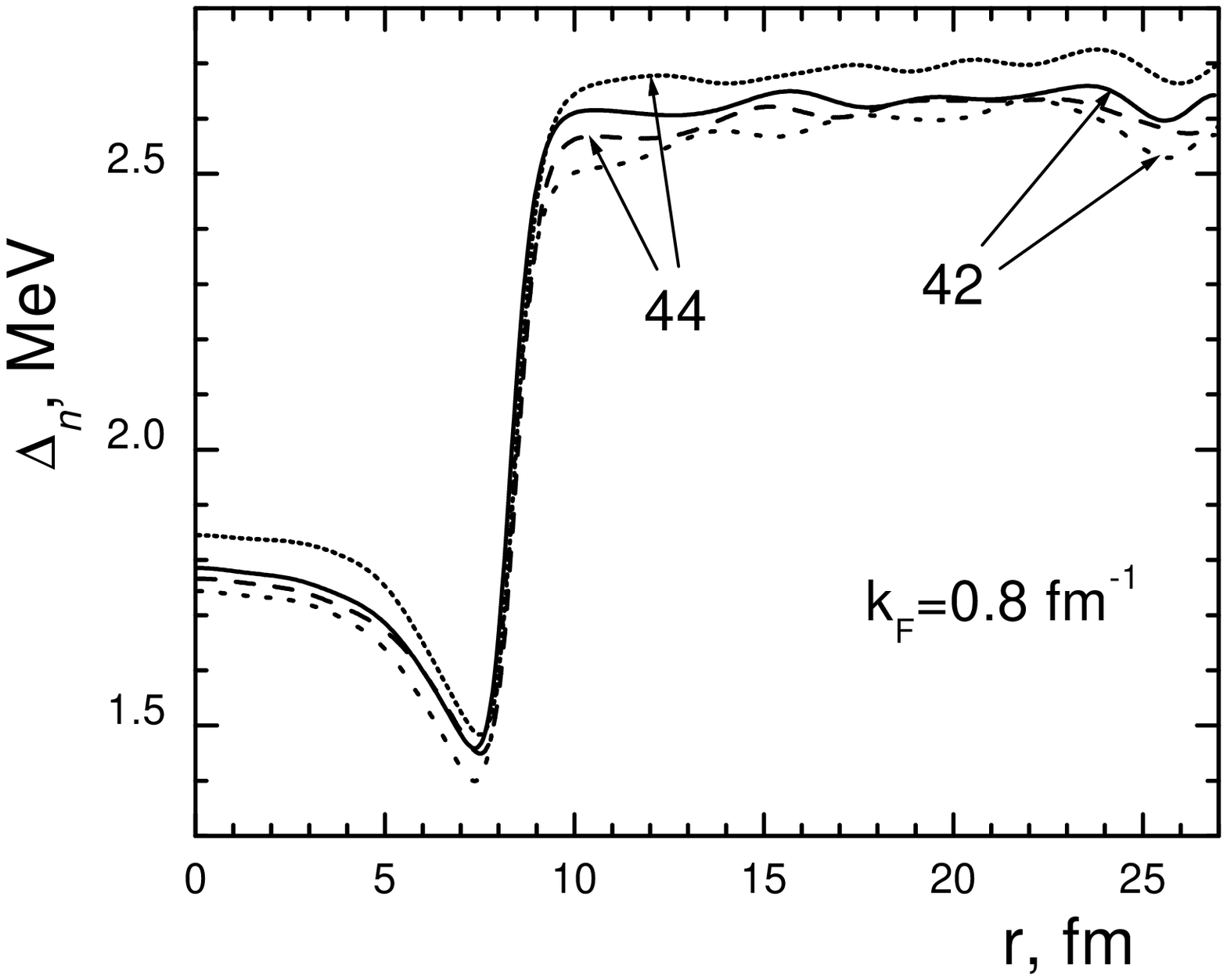}}%
\vspace{0mm} \caption{The neutron gap for
$k_{\Fs}{=}0.8\;$fm$^{-1}$, $Z{=}42$ and $Z{=}44$, in the BC1 case
 (solid lines) and in the BC2 one (dashed lines).}
\end{figure}

 Fig. 4 collects predictions for
$\Delta(r)$ in the BCS approximation for these values of
$k_{\Fs}$. In accordance with the above agreement, the BC1 kind of
the boundary conditions is used. The value of
$k_{\Fs}{=}1\;$fm$^{-1}$ is included as an optional one as far as
in this case the uncertainty is not negligible (about 20\%) but it
is not so big as the one for higher values of $k_{\Fs}$.

\begin{figure}
\centerline{\includegraphics [height=100mm,width=120mm]{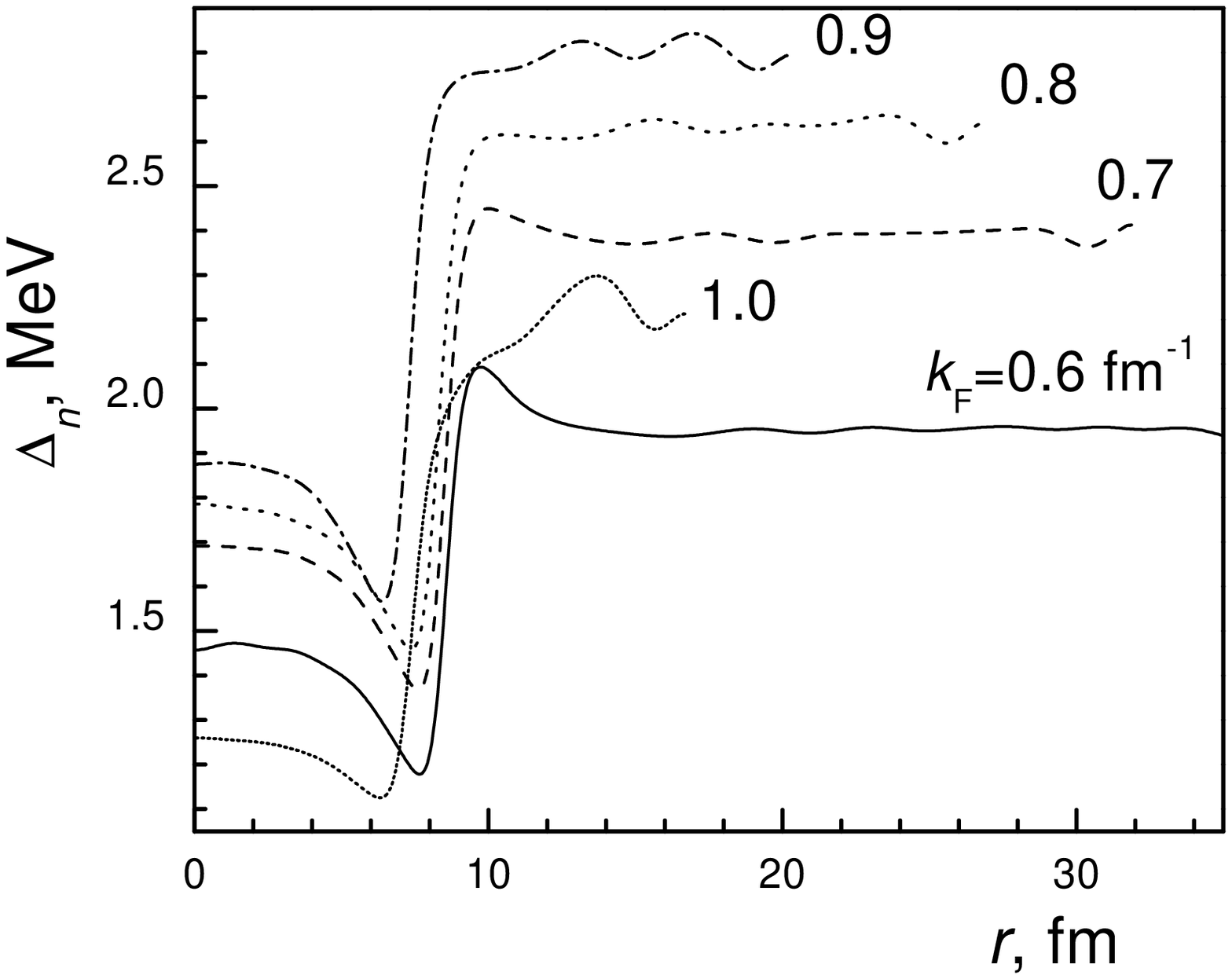}}
\vspace{2mm} \caption{ The gap $\Delta(r)$
 for $k_{\Fs}{=}0.6 \div 1.0  \;$fm$^{-1}$, the P1 (BCS) model, and the BC1 version
 of boundary condition.}
\end{figure}

On the contrary, as it is seen in Table 2, in the case of high
densities, $k_{\Fs} \gsim 1\;$fm$^{-1}$, the uncertainty in the
value of the neutron gap is rather large. As it was shown in
\cite{BC12}, such big variations (BC1 {\it versus} BC2) of the gap
value in the WS approximation appear due to the shell effect in
the neutron single-particle spectrum which is rather pronounced in
the case of high $k_{\Fs}$ and, correspondingly, small  $R_c$
values. We consider this effect as an artifact of the WS method
which should disappear in a more consistent approach. In
\cite{BC12}, we suggested an approximate recipe to avoid this
uncertainty for the gap function $\Delta(r)$. This topic will be
discussed in more detail in the next two sections where the P2 and
P3 models are considered with the many-body corrections to the BCS
approximation taken into account.

\section{ Pairing effects in the case of the P2  model}

Let us go to the P2 model in which the many-body corrections to
the BCS approximation are taken into account in an approximate way
outlined in Section 2, with the factor  $f_{\rm m-b}{=}1/2$  in
Eq.~(\ref{fac}) for the gap $\Delta(k_{\Fs})$ which is the input
 for finding the effective pairing interaction from  Eq.~(\ref{DeltaF}).
The calculation scheme itself and the presentation of  results are
mainly similar to those for the P1 model in the previous section.
In particular, again two kinds of boundary conditions, BC1 and
BC2, are used, and the dependence of the results on the choice is
analyzed. Fig.~5 shows the values of the binding energy per
nucleon for the P2 model, to be compared to Fig.~2, where the P1
model is considered. It is not so detailed as Fig.~2 because,
according to the experience within the P1 model, we limit the
analysis to the vicinity of the absolute minimum of the function
$E_{\rm B}(Z)$ as found for the P1 model. Let us first discuss the
results with the BC1 boundary conditions. In this case, a detailed
analysis was made for $k_{\Fs}{=} 0.9\;$fm$^{-1}$, $k_{\Fs}{=}
1.1\;$fm$^{-1}$ and $k_{\Fs}{=} 1.2\;$fm$^{-1}$. Comparison with
Fig.~2 shows that, at a fixed value of $k_{\Fs}$, the positions of
the local minima in the P2 model are close to those in the P1
model. Besides, the relative positions of the absolute minimum and
the local ones are the same for the two models. Therefore for
other values of $k_{\Fs}$ we examined only the vicinity of the
corresponding absolute minimum in the P1 model. The only exception
is the ``suspicious'' case of $k_{\Fs}{=} 0.7\;$fm$^{-1}$ for
which, in the P1 model, the values of $E_{\rm B}(Z)$ in two local
minima are rather close. The systematic calculation (with the step
$\delta Z{=}4$, instead of the regular one, $\delta Z{=}2$) showed
that in this case the general rule formulated above is also valid.

Let us consider now the results in the case of the BC2 kind of
boundary conditions. Again the situation is quite similar to that in
the P1 model, as confirmed by the detailed comparison of the two
sets of calculations for the BC1 and BC2 versions in the case of
$k_{\Fs}{=} 0.9\;$fm$^{-1}$. Therefore, just as in the P1 model, in
the case of the BC2 boundary conditions one can limit the analysis
in the vicinity of the absolute minimum for the BC1 one.

 Table 3 presents the main ground state characteristics of the
 neutron star inner crust within the P2 model. It is organized
 similarly to Table 1. One can see that for all the values of
$k_{\Fs}$ under consideration, with the exception of $k_{\Fs}{=}
0.8\;$fm$^{-1}$, the equilibrium values of $Z$ in the BC1 and BC2
cases are different. Let us now compare the  equilibrium $Z$ values
in the P2 model with those in the P1 model (Table 1) for the BC1
case. One can see that at high densities, $k_{\Fs} \ge
1.0\;$fm$^{-1}$, they coincide, being equal to $Z{=}20$. The maximal
difference, $\delta Z{=}Z^{(\rm P1)}{-}Z^{(\rm P2)}{=}6$, occurs in
the case of $k_{\Fs}{=} 0.7\;$fm$^{-1}$. It can be explained with
the very flat dependence of $E_{\rm B}$ on $Z$ in this case.
Therefore any change of the calculation parameters, namely, of the
gap value in neutron matter $\Delta_n$ in the P2 model versus the P1
one, can shift the position of the minimum significantly. The
difference of the WS cell radius values, $\delta R_c{=}R_c^{(\rm
P1)}{-}R_c^{(\rm P2)}$, at a given $k_{\Fs}$ appears mainly due to
the difference in $\delta Z$. It is usually of the order of 1 fm,
and only in the same case of $k_{\Fs}{=} 0.7\;$fm$^{-1}$ $\delta
R_c$ exceeds 2 fm. Let us now analyze the influence of the kind of
boundary conditions on the ground state characteristics in the case
of the P2 model.  In general, this effect is of the same magnitude
as in the P1 model. Again the uncertainty in the equilibrium value
of $Z$ is between 2 and
 6 units and in the value of $R_c$,  about $1 - 2\;$fm.
A more detailed comparison with Table 1 shows that, as a rule, the
effect of the boundary conditions in the P2 model is less than that
in the P1 one, but only a little. For example, at $k_{\Fs}{=}
0.8\;$fm$^{-1}$ the equilibrium Z values for the BC1 and BC2 cases
are now equal to each other. However, there is the case of
$k_{\Fs}{=} 0.9\;$fm$^{-1}$ in which the effect under discussion is
stronger in the P2 model. It should be noted that in the case of
$k_{\Fs}{=} 1.2\;$fm$^{-1}$ the first line of the table ($Z{=}20$)
contains empty positions corresponding to the BC2 version. This
means that in the case under consideration (the P2 model, the BC2
type of boundary conditions and  $Z{=}20$) the WS solution of the
type we consider, i.e. the WS cell with a nuclear type cluster in
the center, is not stable. This is a signal of proximity to the
point of instability for the phase transition to the homogeneous
state. In fact, for such high density values corresponding to
$k_{\Fs}{=} 1.2\;$fm$^{-1}$ \cite{Mag,Oyam} and, maybe, to
$k_{\Fs}{=} 1.1\;$fm$^{-1}$ \cite{Mag} the so-called ``spaguetti''
phase should appear which can not be described within the WS method
with the spherical symmetry assumed. Therefore our consideration of
these $k_{\Fs}$ should be considered as optional, and the
corresponding results are reported mainly for methodological reason.

\begin{table}
\caption{Comparison of characteristics  of equilibrium
configurations of the WS cell for two different kinds of the
boundary conditions in the case of the P2 model.}
\bigskip
\begin{center}
\begin{tabular}{|c|c|c|c|c|c|c|c|}
\hline
  {$k_{\rm F,}$} & \raisebox{-6pt}{$\,\,Z\,$} &
  \multicolumn{2}{|c|}{$R_{\rm c},\;$fm}\rule{0pt}{14pt}&
  \multicolumn{2}{|c|}{$E_{\rm B},\;$MeV}&
  \multicolumn{2}{|c|}{$\mu_{\rm n},\;$MeV}\\
\cline{3-8}
 \rule{0pt}{13pt}${\rm fm^{-1}}$ && BC1 & BC2 & BC1 & BC2 & BC1 & BC2 \\
\hline
 \raisebox{-6pt}{ 0.6} & 56 & 36.85 & 36.88 & 2.2837 & 2.2842 & 3.4755
  & 3.4818 \\
      & 54 & 36.02 & 36.04 & 2.2838 & 2.2839 & 3.4660 & 3.4629    \\
\hline
  \raisebox{-6pt}{0.7} & 46 & 30.31 & 30.28 & 2.9320 & 2.9354 & 4.2892
   & 4.2451 \\
      & 48 & 30.98 & 30.88 & 2.9332 & 2.9348 & 4.3004 & 4.2533   \\
\hline\rule{0pt}{14pt}
  0.8 & 40 & 25.97 & 26.19 & 3.5781 & 3.5807 & 5.0762 & 5.0987 \\
\hline
  \raisebox{-6pt}{0.9} & 20 & 18.34 & 18.60 & 4.2173 & 4.2452 & 5.7027
   & 5.8477 \\
      & 26 & 20.38 & 20.93 & 4.2416 & 4.2273 & 5.8894 & 6.0839   \\
\hline
  \raisebox{-6pt}{1.0} & 20 & 16.56 & 16.94 & 4.8641 & 4.9427 & 7.5247
   & 6.8272 \\
      & 24 & 18.25 & 17.80 & 4.9411 & 4.9013 & 7.0171 & 6.3298    \\
\hline
  \raisebox{-6pt}{1.1} & 20 & 15.05 & 15.42  & 5.5734 & 5.7363 & 9.0601
   & 7.9905 \\
      & 24 & 16.56 & 16.16 & 5.7339 & 5.6387 & 8.2971 & 6.9493  \\
\hline
 \raisebox{-6pt}{1.2}  & 20 & 13.73 &   -    & 6.4175 &    -   & 8.1824
 &   -     \\
      & 26 & 15.30 & 14.90 & 6.9244 & 6.4566 & 8.9542 & 10.4879   \\[2pt]
 \hline
\end{tabular}
\end{center}
\end{table}

Let us now turn to the analysis of the neutron gap. The main gap
characteristics in the P2 model are collected in Table 4 which is
analogous to Table 2 in the previous section. Comparison of these
two tables shows that the main effect of changing from the P1 to the
P2 model is quite trivial. It is a general suppression of the gap
characteristics approximately by a factor two in comparison with the
P1 model (i.e. within the BCS approximation). The ratio
$\Delta^{(\rm P2)}/\Delta^{(\rm P1)}$ is equal to 1/2 exactly for
the quantity $\Delta^0_{\rm inf}$ and approximately for $\Delta_{\rm
inf}$. For other characteristics, the deviation of the ratio from
1/2 is usually a little greater than for $\Delta_{\rm inf}$ and only
in few cases it is significant. The Fermi average value $\Delta_{\rm
F}$ is the most important neutron gap characteristic. Let us compare
its values for the BC1 and BC2 versions of the boundary conditions.
Just as in the P1 model, the difference $\Delta_{\rm F}^{\rm BC1}
{-} \Delta_{\rm F}^{\rm BC2} $ is very small for $k_{\Fs}{=} 0.6
\div 0.8 \;$fm$^{-1}$. For $k_{\Fs} > 0.8 \;$fm$^{-1}$ the influence
of the boundary conditions on the $\Delta_{\rm F}$ value becomes
rather strong, significantly stronger than in the P1 model.
Especially strong effect appears for $k_{\Fs}{=} 1.0\;$fm$^{-1}$ and
$k_{\Fs}{=} 1.1\;$fm$^{-1}$ where $\Delta_{\rm F}$ almost vanishes
in the BC1 case, being rather big in the BC2 one. At $k_{\Fs}{=}
1.2\;$fm$^{-1}$, the contrary situation takes place, i.e.
$\Delta_{\rm F}$  vanishes in the BC2 case.

\begin{table}
\caption{Average gap characteristics in the P2 model}
\bigskip
\begin{tabular}{|c|c|c|c|c|c|c|c|c|c|c|c|c|}
\hline\rule{0pt}{14pt}
  {$k_{\rm F},\!\!$} & \raisebox{-6pt}{$Z$} &
  \multicolumn{2}{|c|}{$k_{\rm F}^{\rm as},\,{\rm fm^{-1}}$}&
  \multicolumn{2}{|c|}{$\Delta(0),\;$MeV}&
  \multicolumn{2}{|c|}{$\Delta_{\rm as},\;$MeV}&
  \multicolumn{2}{|c|}{$\Delta_{\rm F},\;$MeV}&
  \multicolumn{2}{|c|}{$\Delta_{\rm inf},\;$MeV}&
  {$\Delta^0_{\rm inf},\!\!$}\\
\cline{3-12}
 \rule{0pt}{13pt}$\rm fm^{-1}\!\!$&& BC1 & BC2 & BC1 & BC2 & BC1 & BC2 & BC1 & BC2 & BC1 &  BC2
 &MeV$\!\!$\\
 \hline\rule{0pt}{13pt} \raisebox{-6pt}{0.6}& 56 &0.5797 &0.5790 &0.971 & 0.962 & 1.058 & 1.056 &1.059
&1.051 & 1.163 &1.161 & \raisebox{-6pt}{1.21} \\
 & 54 & 0.5783 &0.5781 &0.980 &0.946 &1.063 &1.042
&1.050 &1.048 &1.160 &1.159 &\\

\hline \raisebox{-6pt}{0.7} & 46 &0.6760 &0.6760 &1.057 &1.104
&1.241 &1.244
&1.211 &1.244 &1.340 &1.341 &\raisebox{-6pt}{1.38} \\
& 48 & 0.6784 & 0.6750 &1.141 &1.044 &1.248 &1.202 &1.235 &1.213 &1.345 &1.339 &  \\
\hline\rule{0pt}{13pt} {0.8} &40 &0.7691 & 0.7752 & 1.025 &1.231
&1.264 &1.360
&1.256 &1.346  &1.438 &1.443 &{1.46}\\
%& 44 &0.7729 &0.7727 &1.747 &1.834 &2.580 &2.679 &2.525 &2.560 &2.883 &2.883 & \\
\hline \raisebox{-6pt}{0.9} &20 & 0.8514 &0.8769 &0.608 &1.271
&0.834 &1.374
&0.816 &1.401 &1.460 &1.459 &\raisebox{-6pt}{1.46}\\
& 26 &0.8670 &0.8820 &1.113 &1.325 &1.305 &1.323 &1.320 &1.320 &1.459 &1.459 & \\
\hline \raisebox{-6pt}{1.0} &20 &0.9399 &0.9675 &0.024 &1.523
&0.041 &1.329
&0.037 &1.453 &1.411 &1.380 & \raisebox{-6pt}{1.34} \\
& 24 &0.9630 &0.9344 &1.465 &0.447 &1.339 &0.644 &1.426 &0.621 &1.385 &1.418 &\\
\hline \raisebox{-6pt}{1.1} &20 &1.0315 &1.0580 &0.015 &1.576
&0.026 &1.245
&0.022 & 1.424 &1.275 &1.219 &\raisebox{-6pt}{1.13}\\
& 24 &1.0549 &1.0258 &1.439 &0.804 &1.253 &0.649 &1.393 &0.658 &1.226 &1.287 & \\
\hline \raisebox{-6pt}{1.2} &20 &1.1253 &- &1.229 &-
 &0.461 &-
&0.539 &- &1.053 &- &\raisebox{-6pt}{0.83}\\
& 26 &1.1306 &1.1146 &0.177 &0.050 &0.305 & 0.068 &0.309 &0.061
&1.037 &1.086&\\[2pt]
 \hline
\end{tabular}
\end{table}

\begin{figure}
\includegraphics [height=180mm,width=120mm]{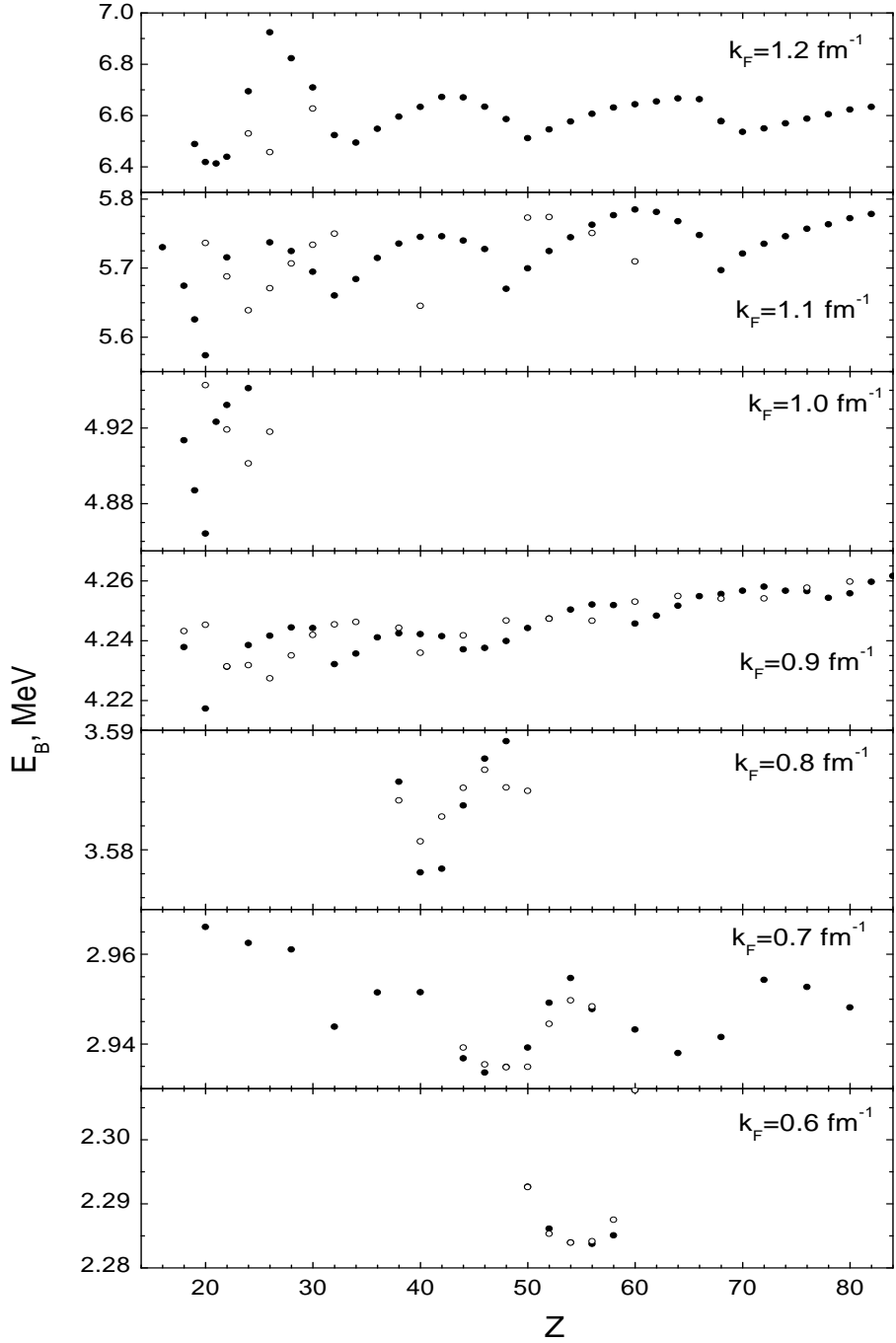}%
\vspace{5mm} \caption{ Binding energy per a nucleon for various
$k_{\Fs}$ within the P2 model  in the case of the BC1 (solid
circles) and the BC2 (open  circles) kinds of the boundary
conditions.}
\end{figure}

For the case of $k_{\Fs}{=} 1.1\;$fm$^{-1}$, the $\Delta(r)$
function is displayed in Fig. 6 for the two kinds of boundary
conditions.  The equilibrium value is $Z{=}20$ in the BC1 case and
$Z{=}24$ in the BC2 one. As one can see, the difference between
predictions of the two kinds of boundary conditions is drastic. The
most strong variation of the gap occurs in the case of $Z{=}20$. To
understand the reason of this effect, it is instructive to examine
the neutron single particle spectrum $\eps_{\lambda}$. It is
displayed  in Fig. 7 for the BC1 version and the BC2 one. The
positions of the chemical potential $\mu_n$ are shown with dots. The
two spectra are essentially different. The reason is the shift
$\delta \eps_{\lambda}$  of each $\lambda$-level going from BC1 to
BC2 version. The value of the shift is approximately equal to one
half of the distance between two neighboring levels with the same
($l,j$),
 the sign of the shift being opposite for even and odd $l$. The
absolute value of the shift is proportional to $1/R_c^2$ and grows
at increasing values of $k_{\Fs}$. These  shifts are shown  for two
states, 2$j_{13/2}$ and 1$n_{23/2}$, which are the neighbors of
$\mu_n$ in the BC1 case. One can see that in both cases there is a
shell type structure with rather wide intervals between some
neighboring levels. In the BC1 case, we deal with a big inter-level
space just at Fermi surface, $\mu_n$ being inside. The width of this
interval exceeds the value of $2\Delta_{\rm inf} \simeq 2.5\;$MeV
which is characteristic for the gap equation. That is why the gap
equation in the BC1 case has practically zero solution ($\Delta_{\rm
F} << \Delta_{\rm inf}$). In the BC2 case, big intervals are
situated far from the Fermi surface and do not influence
significantly the gap equation. Therefore  we have a normal
situation in this case with $\Delta_{\rm F} \simeq \Delta_{\rm
inf}$. For $Z{=}24$ which is the equilibrium value for the BC2 case,
the difference between predictions of the BC1 and BC2 versions is
not so dramatic but also exists.

Let us return to the P1 model (Table 2). One can see that in the
case of  $k_{\Fs}{=} 1.1\;$fm$^{-1}$ and $Z{=}20$ the Fermi
average gap value $\Delta_{\rm F}$ is suppressed in comparison
with the normal one, but is not zero. For the solution of the gap
equation one can use the same spectrum $\eps_{\lambda}$ displayed
in Fig. 7. Indeed, the only difference between P1 and P2 models at
a fixed value of $Z$ is the value of the gap in neutron matter,
$\Delta^0_{\rm inf}$, which is a parameter of the model. In the P1
model it is about two times larger than that in the P2 model.
However, the direct influence of the gap on the mean field
potential and the single-particle spectrum is negligible. But now
the value $2\Delta_{\rm inf} \simeq 5\;$MeV is of the order of the
energy interval under discussion. Therefore the suppression effect
in the gap equation is less than the one in the P2 model.

The ``Shell effect'' in the neutron single-particle spectrum was
discussed in detail in \cite{BC12}. It was interpreted there as an
artifact of the WS approach which does not take into account the
periodicity of the crystal. It should disappear in a more consistent
approach to the neutron star inner crust structure with periodical
boundary conditions. A recipe was suggested in \cite{BC12} for
improving this drawback  and finding the gap $\Delta(r)$ in such
anomalous cases with big difference of the gap values in the BC1 and
BC2 versions, one of them being strongly suppressed. It is based on
the smooth dependence of $\Delta(r)$ on $Z$ in a regular situation
and consists in the use of $\Delta(r)$ for a neighboring $Z$ with a
regular single-particle spectrum at the Fermi surface. In the case
under consideration, Fig. 6, the solid line (BC1) for $Z{=}24$ or
the dashed one (BC2) for $Z{=}20$ correspond to the ``normal''
situation. The difference between these two curves is not greater
than 10\%, and, within such accuracy, any of them could be used as
the prediction for $\Delta(r)$ in the case of
$k_{\Fs}{=}1.1\;$fm$^{-1}$ within the P2 model. As it was discussed
in \cite{BC12}, such a recipe is not self-consistent within the WS
method but it seems to be reasonable from the physical point of
view. The arguments were given in this article in favor of the
conclusion that the solution of the gap equation in a more
consistent approach should be close to the one in the WS
approximation in the case of the regular situation for the neutron
spectrum.

\begin{figure}
\centerline{\includegraphics [height=100mm,width=120mm]{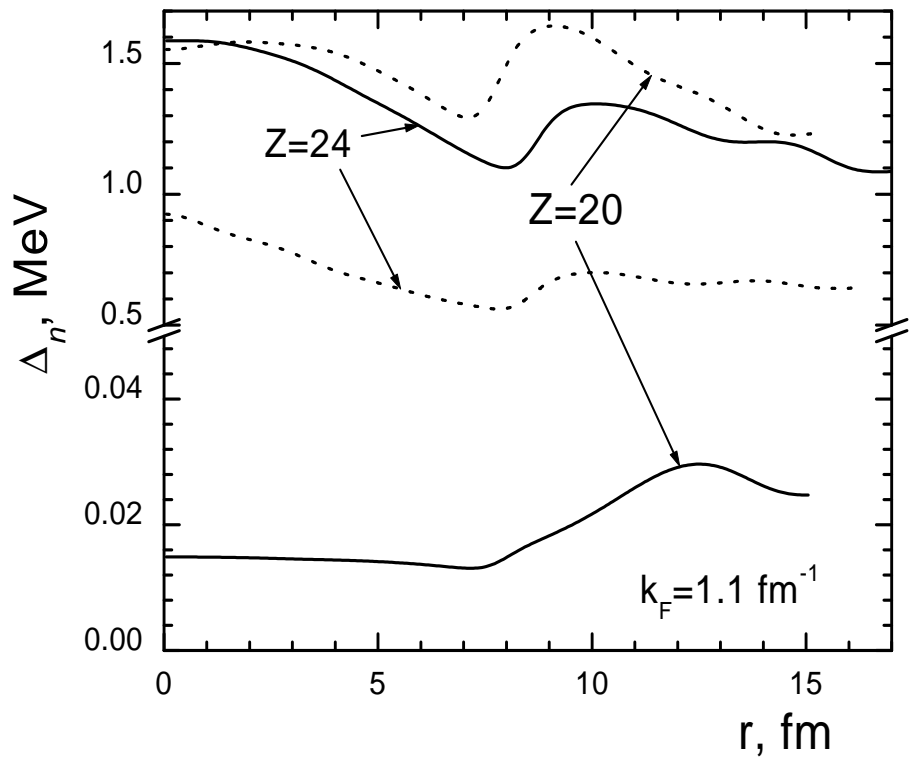}}
\vspace{2mm} \caption{  The neutron gap function $\Delta_n(r)$ for
$k_{\Fs}{=}1.1\;$fm$^{-1}$, $Z{=}20$ and $Z{=}24$, in the BC1 case
 (solid lines) and in the BC2 one (dots), within the P2 model.}
\end{figure}

\begin{figure}
\centerline{\includegraphics [height=130mm,width=100mm]{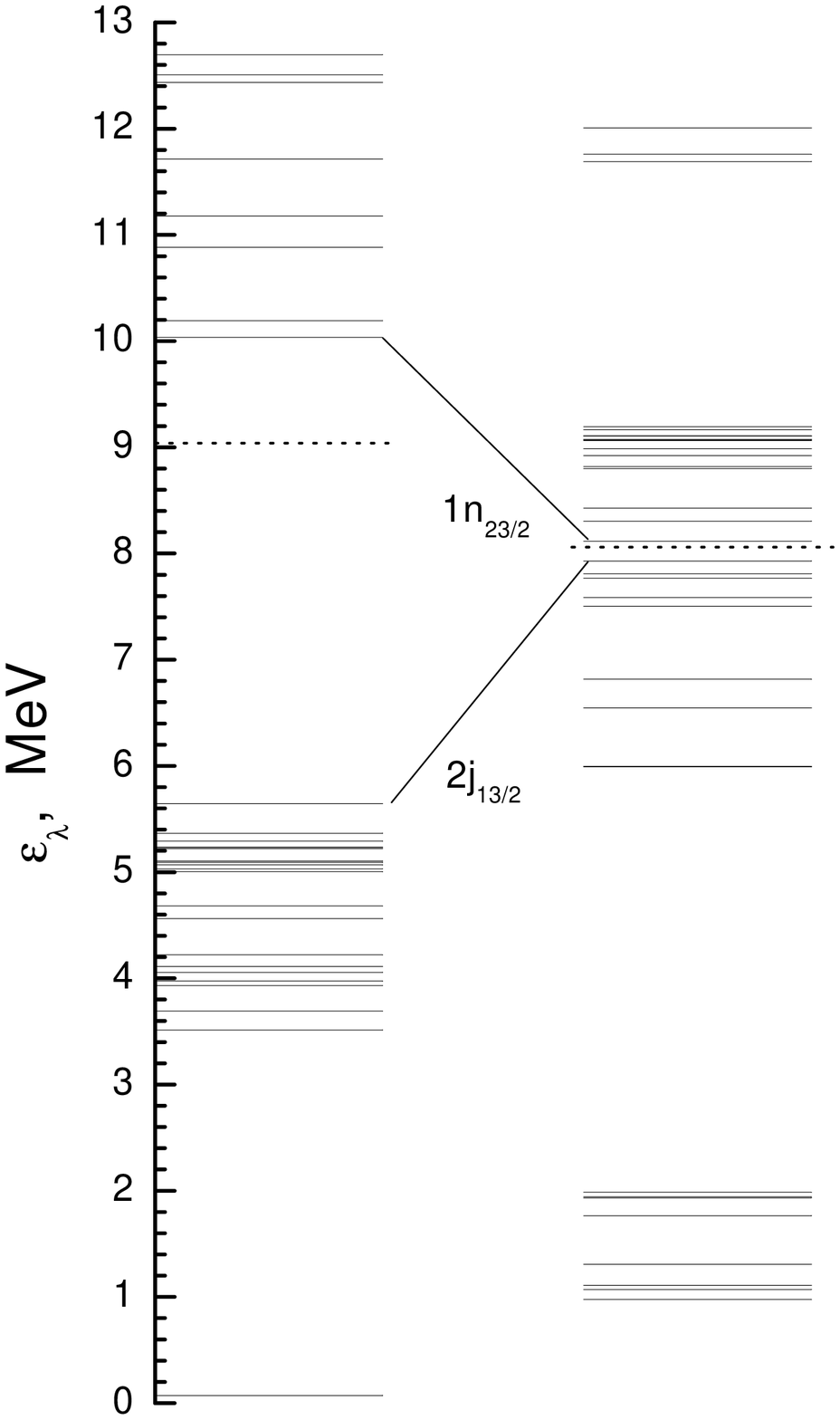}}
\vspace{5mm} \caption{ The neutron single-particle spectrum
$\eps_{\lambda}$ for $k_{\Fs}{=}1.1\;$fm$^{-1}$, $Z{=}20$, in the
 BC1 case (left) and the BC2 one (right), for the P2 model.}
\end{figure}

As it is seen in Table 4, at intermediate densities, $k_{\Fs}{=}0.6
\div 0.8 \;$fm$^{-1}$, the regular situation takes place with the
approximate equality $\Delta_{\rm F} \simeq \Delta_{\rm inf}$ for
both kinds of boundary conditions and $\Delta_{\rm F}^{\rm BC1}
\simeq \Delta_{\rm F}^{\rm BC2} $. Therefore one can expect that the
gap function $\Delta_n(r)$ will be approximately the same in the BC1
and BC2 cases. At $k_{\Fs}{=}0.8\;$fm$^{-1}$, where the equilibrium
$Z$ value is the same in both versions, these functions are
displayed in Fig. 8. In this case the difference between
$\Delta_{\rm F}^{\rm BC1}$ and $\Delta_{\rm F}^{\rm BC2}$ is about
10\%. As it can be seen in Fig. 8, the difference $\Delta_n^{\rm
BC1}(r) - \Delta_n^{\rm BC2}(r)$ is also about 10\% with the
exception of small $r<3\;$fm which doesn't contribute appreciably to
the matrix elements of $\Delta_n$. Thus, the accuracy of predictions
for the gap function within the WS approach in the P2 model for the
density interval under consideration is about 10\%. Fig. 9 collects
the predictions for the gap functions for these values of $k_{\Fs}$
which are taken in accordance with the prescription of Section 3,
i.e. those for the BC1 kind of boundary conditions.

\begin{figure}
\centerline{\includegraphics [height=100mm,width=120mm]{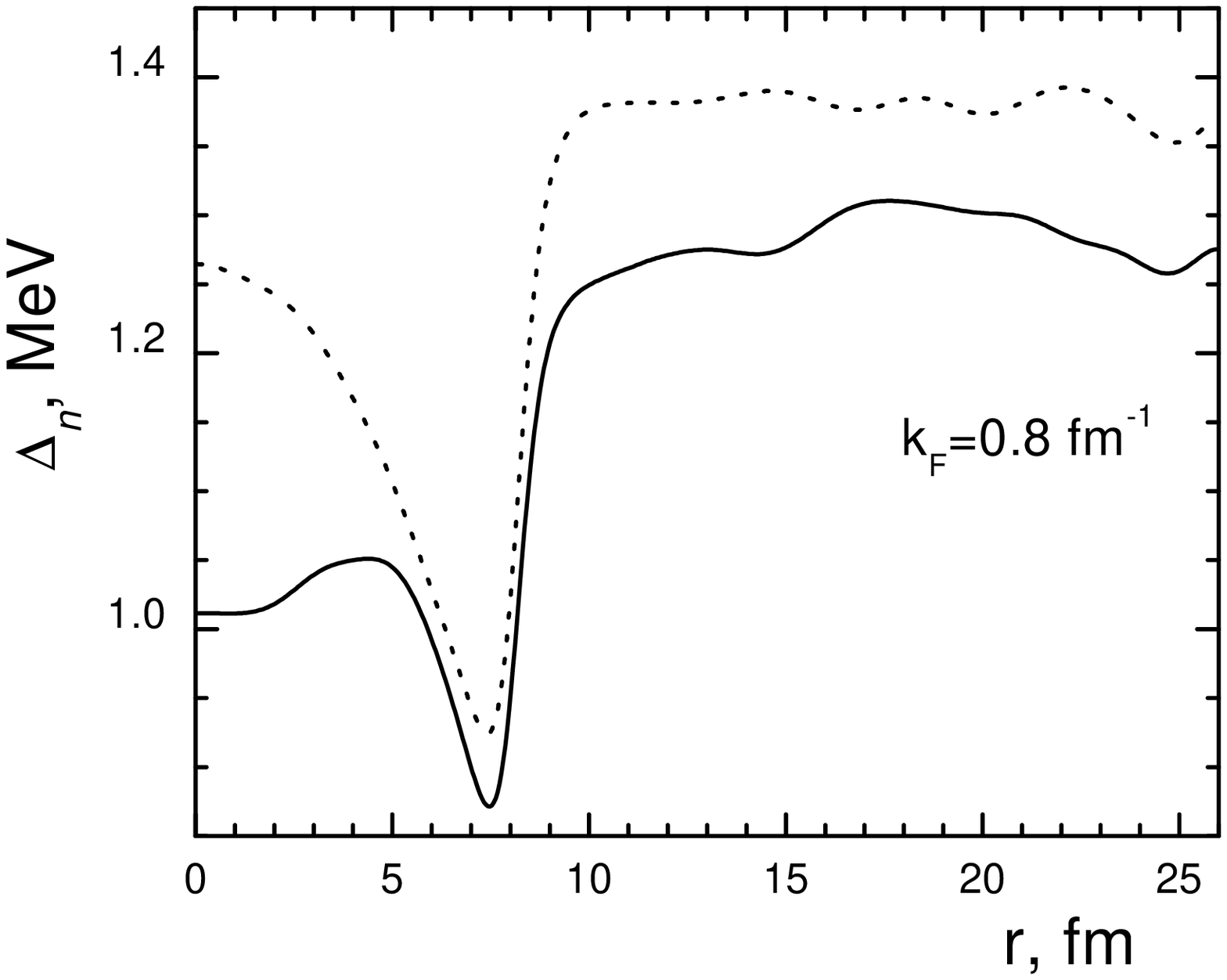}}
\vspace{2mm} \caption{ The neutron gap function $\Delta_n(r)$ for
$k_{\Fs}{=}0.8\;$fm$^{-1}$, $Z{=}40$, for the P2 model in the BC1
case (solid line) and in the BC2 one (dashed line).}
\end{figure}

\begin{figure}
\centerline{\includegraphics [height=100mm,width=120mm]{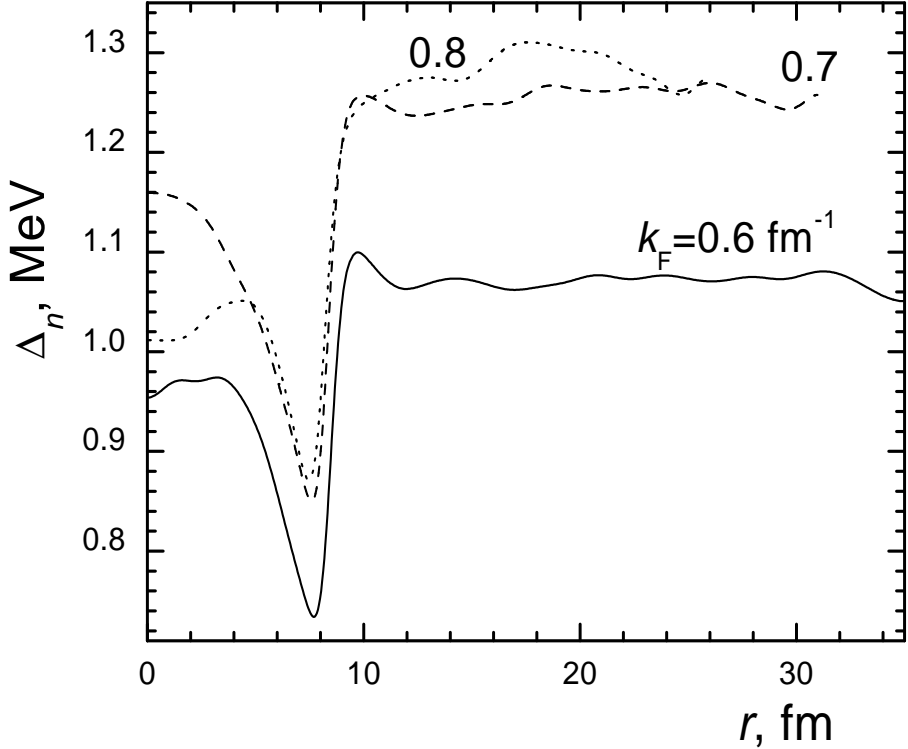}}
\vspace{2mm} \caption{ The neutron gap $\Delta_n(r)$
 for $k_{\Fs}{=}0.6 \div 0.8  \;$fm$^{-1}$ in the case
of the P2 model.}
\end{figure}

Let us go to higher density values, $k_{\Fs} \ge 0.9\;$fm$^{-1}$,
where the difference between the self-consistent values of
$\Delta_{\rm F}^{\rm BC1}$ and $\Delta_{\rm F}^{\rm BC2}$ is
significant. As the above discussion for $k_{\Fs}{=}1.1
\;$fm$^{-1}$ showed, there are two possibilities for the choice of
$\Delta_n(r)$ in this case with close results. To be definite, let
us use the BC1 version as the basic one if the relation
$\Delta_{\rm F}^{\rm BC1} \simeq \Delta_{\rm inf}$ takes
place.\footnote {There is a tiny difference between the values of
$\Delta_{\rm inf}$ in the two versions of the boundary conditions
which originates from a small difference of the two values of the
asymptotic density. However, it is much less than the effects
under discussion.} In the opposite case, $\Delta_{\rm F}^{\rm BC1}
<< \Delta_{\rm inf}$, we use the equilibrium $Z^{\rm BC1}$ value
and the gap function $\Delta_n^{\rm BC2}(r)$ for this $Z$.
Corresponding gap functions are displayed in Fig. 10. The analysis
similar to that for $k_{\Fs}{=}1.1 \;$fm$^{-1}$ shows that the
accuracy of the predictions for $\Delta_n(r)$ is again about 10\%
and only for $k_{\Fs}{=}1.2 \;$fm$^{-1}$ it is a little worse.

\newpage
\begin{figure}
\centerline{\includegraphics [height=100mm,width=120mm]{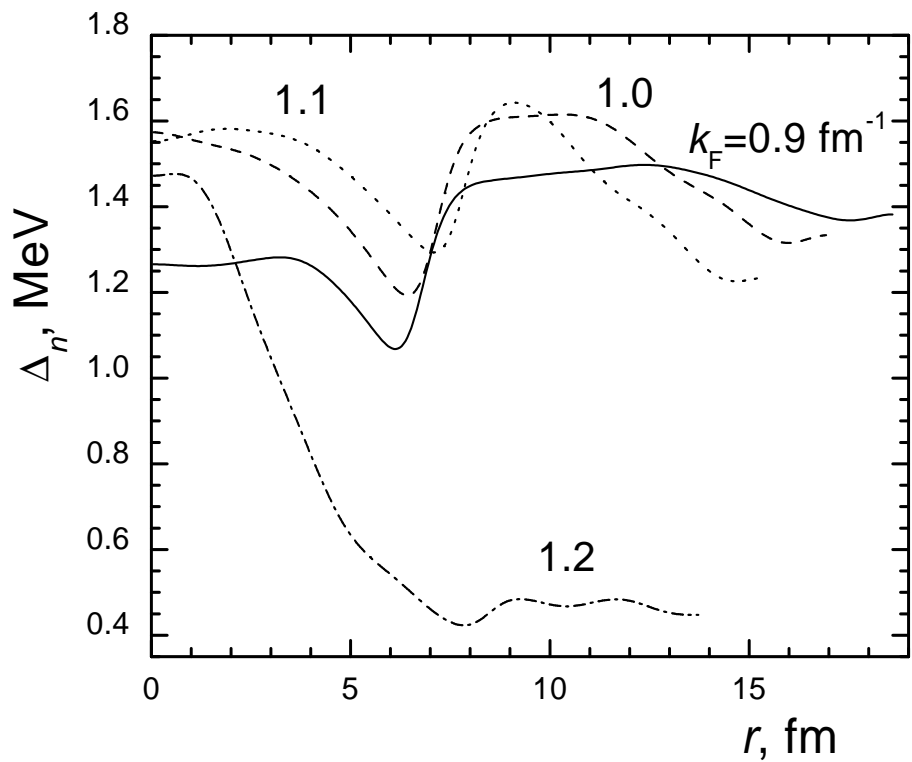}}
\vspace{2mm} \caption{ The neutron gap $\Delta_n(r)$
 for $k_{\Fs}{=}0.9 \div 1.2  \;$fm$^{-1}$ in the case
of the P2 model.}
\end{figure}

\section{The P3  model}

The P3 model  is similar to the P2 one, but now  the many-body
suppression factor  in Eq.~(\ref{fac}) is equal to $f_{\rm
m-b}{=}1/3$. The calculation scheme and the presentation of results
is quite similar to that in the previous section. Values of the
binding energy per a nucleon $E_{\rm B}$ in the P3 model are given
in Fig.~11 for different densities, $k_{\Fs}{=}0.6 \div 1.2
\;$fm$^{-1}$, similarly to Fig.~5. Again the detailed analysis is
made for $k_{\Fs}{=}0.9 \;$fm$^{-1}$ and partially  for
$k_{\Fs}{=}0.7 \;$fm$^{-1}$. Comparison with Fig. 5 shows that
positions of the absolute minima of the function $E_{\rm B}(Z)$ are
the same as in the P2 model and in the P3 one for both kinds of
boundary conditions at all  $k_{\Fs}$ values under consideration
with only one exception, $k_{\Fs}{=}0.9 \;$fm$^{-1}$ and the BC2
version. In the latter case the equilibrium $Z$ values differ by 2
units. The main ground state characteristics of the neutron star
inner crust within the P3 model are presented in Table 5 which is
similar to Table 3 in Section 4 or Table 1 in Section 3. As one can
see, the differences between all the values in Table 5 and those in
Table 3 are quite small.

\begin{figure}
\centerline{\includegraphics [height=180mm,width=100mm]{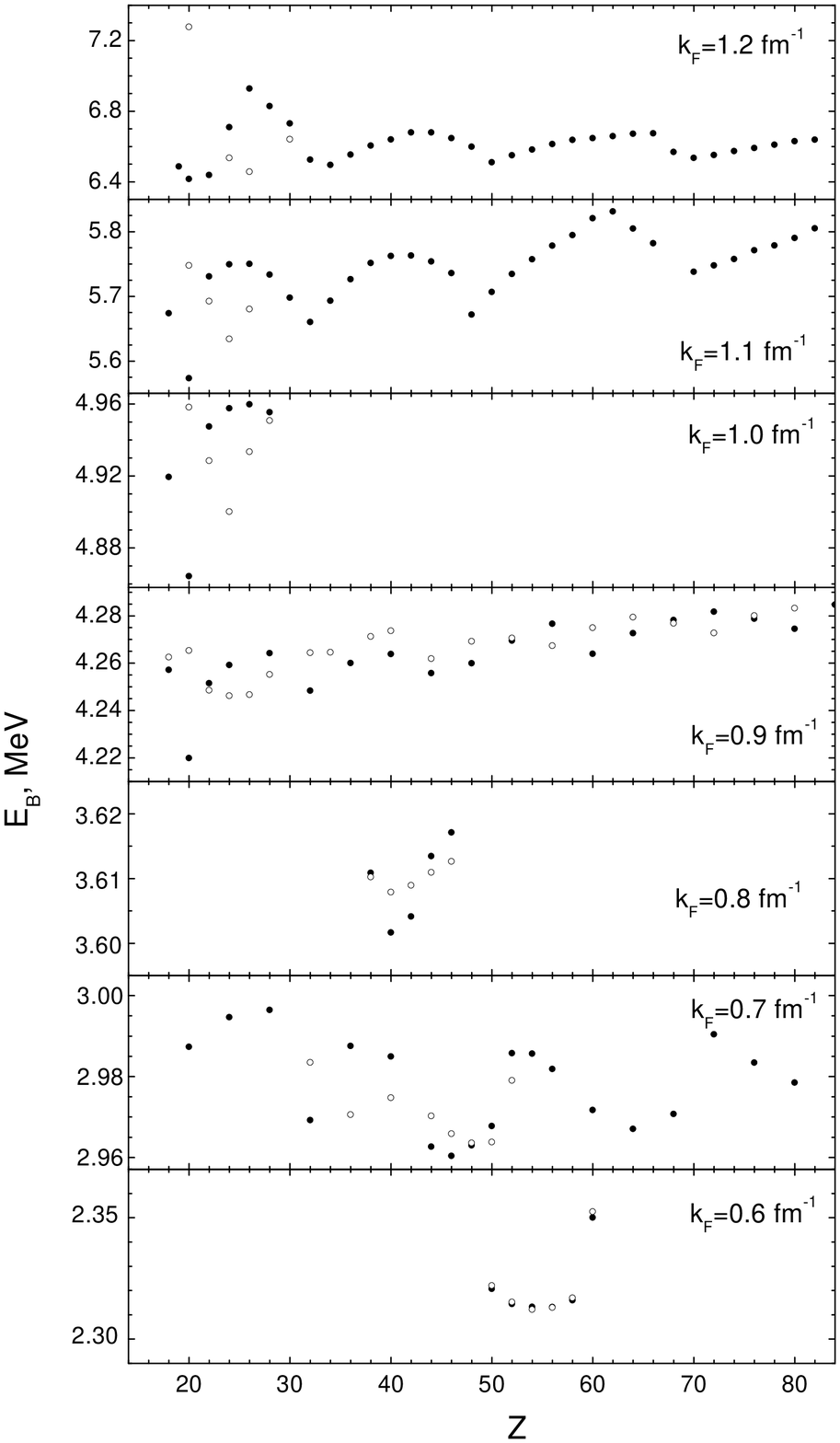}}
\vspace{5mm} \caption{Binding energy per a nucleon for various
$k_{\Fs}$ for the P3 model  in the case of the BC1 (solid circles)
and the BC2 (open  circles) kinds of the boundary conditions.}
\end{figure}

Let us now turn to the analysis of the neutron gap in the P3
model. The corresponding  gap characteristics  are presented in
Table 6 which is similar to Table 4 (the P2 model) or Table 2 (the
P1 model). Comparison with Table 4 shows that again the main
effect is the general decrease of all the gap characteristics by
the factor of 2/3 which corresponds to the values of the many-body
suppression factor $f_{\rm m-b}{=}1/3$ in Eq.~(\ref{fac})  in the
P3 model and $f_{\rm m-b}{=}1/2$ in the P2 model. In  two previous
sections  we discussed a pseudo effect of strong suppression of
the neutron gap in some cases with high values of $k_{\Fs} \gsim
1\;$fm$^{-1}$. It originates due to non-regular behavior at big
$k_{\Fs}$ of the neutron single particle spectra (the ``pseudo
Shell effect'') in the WS approach. In the P1 model (Section 3)
the first case of a moderate suppression occurs at $k_{\Fs}
{=}1.0\;$fm$^{-1}$, in the P2 model (Section 4), at $k_{\Fs}
{=}0.9\;$fm$^{-1}$. In addition, in the P2 model, at $k_{\Fs} \ge
1.0\;$fm$^{-1}$, the gap almost vanishes  in some cases for the
BC1 case or the BC2 one. Table 6 shows that in the P3 model this
pseudo effect becomes even stronger, namely, the first case of
vanishing occurs at $k_{\Fs} {=}0.9\;$fm$^{-1}$. A typical case of
such  vanishing is shown in Fig. 12. The way to improve this
drawback of the WS method and to find the neutron gap function
$\Delta_n(r)$ in such ``bad'' cases in the P3 model is the same as
above. The predictions for $\Delta_n(r)$ within the P3 model are
displayed in Fig. 13 for
 $k_{\Fs}{=}0.6 \div 0.8 \;$fm$^{-1}$ and in Fig. 14 for
 $k_{\Fs}{=}0.6 \div 1.2 \;$fm$^{-1}$. The method to choose the
 version (BC1 or BC2) at every value of $k_{\Fs}$ is the same as it was
 suggested  in the previous section for the P2 model.

\begin{figure}
\centerline{\includegraphics [height=100mm,width=120mm]{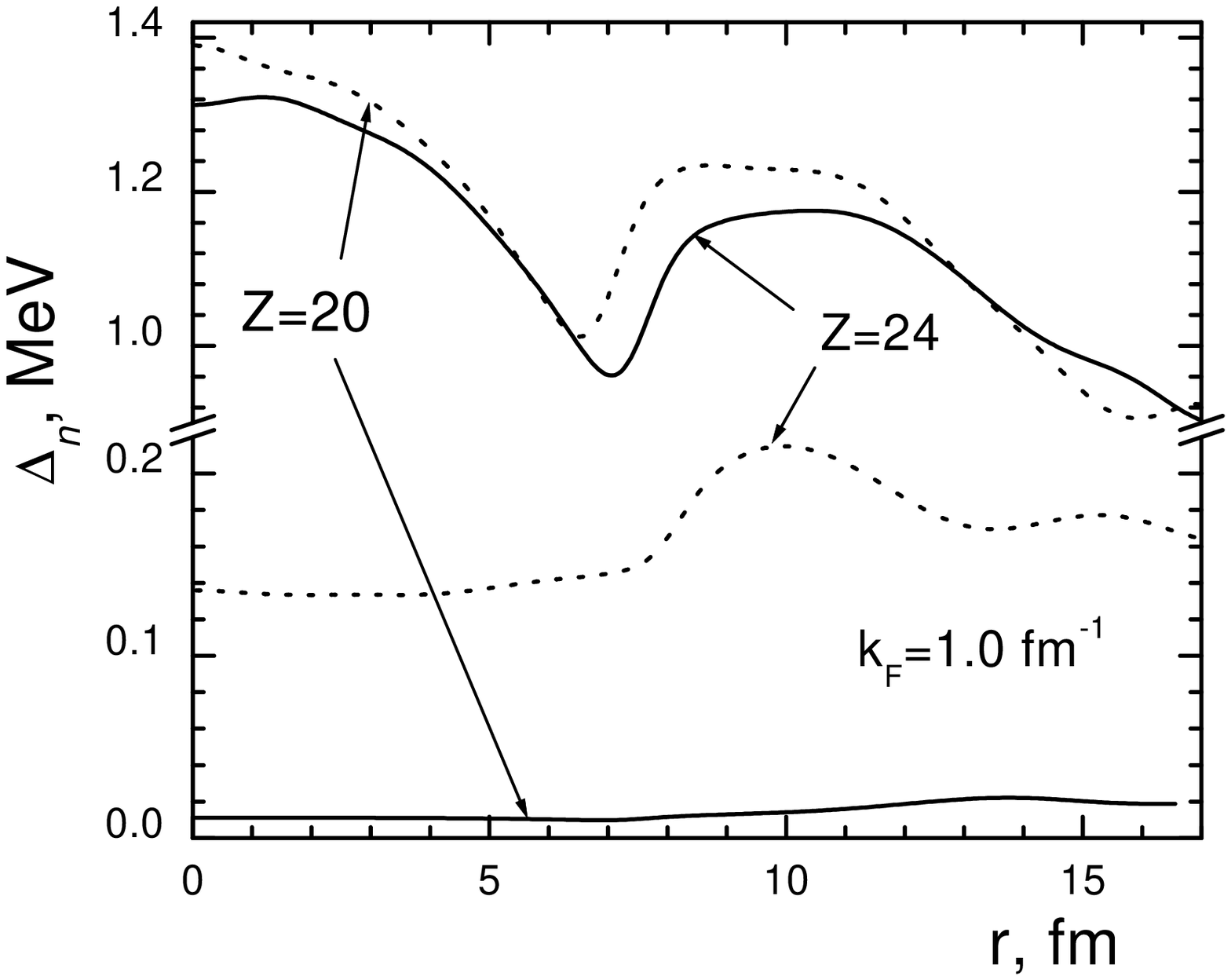}}
\vspace{2mm} \caption{ The neutron gap for
$k_{\Fs}{=}1.0\;$fm$^{-1}$, $Z{=}20$ and $Z{=}24$, in the BC1 case
 (solid lines) and in the BC2 one (dots), for the P3 model.}
\end{figure}

\begin{table} \caption{Comparison of characteristics of equilibrium
configurations of the WS cell for two different kinds of the
boundary conditions in the case of the P3 model.}
\bigskip
\begin{center}
\begin{tabular}{|c|c|c|c|c|c|c|c|}
\hline
  {$k_{\rm F},$} & \raisebox{-6pt}{$\,\,Z\,$} &
  \multicolumn{2}{|c|}{$R_{\rm c},\;$fm}\rule{0pt}{14pt}&
  \multicolumn{2}{|c|}{$E_{\rm B},\;$MeV}&
  \multicolumn{2}{|c|}{$\mu_{\rm n},\;$MeV}\\
\cline{3-8}\rule{0pt}{14pt}
${\rm fm^{-1}}$&& BC1 & BC2 & BC1 & BC2 & BC1 & BC2 \\
\hline\rule{0pt}{13pt}
  \raisebox{-6pt}{0.6} & 56 & 36.92 & 36.96 & 2.3132 & 2.3130 & 3.5474 & 3.5634\\
      & 54 & 36.04 & 36.06 & 2.3133 & 2.3121 & 3.5472 & 3.5315   \\
\hline
  \raisebox{-6pt}{0.7} & 46 & 30.27 & 30.27 & 2.9603 & 2.9658 & 4.4009 & 4.2937\\
      & 48 & 31.09 & 30.85 & 2.9630 & 2.9636 & 4.3857 & 4.2913    \\
\hline\rule{0pt}{13pt}
  0.8 & 40 & 25.89 & 26.27 & 3.6016 & 3.6079 & 5.1069 & 5.1355 \\
\hline
  \raisebox{-6pt}{0.9} & 20 & 18.40 & 18.62 & 4.2199 & 4.2653 & 6.5184 & 5.8724\\
      & 24 & 20.11 & 19.77 & 4.2592 & 4.2462 & 5.9920 & 5.6863    \\
\hline
  \raisebox{-6pt}{1.0} & 20 & 16.56 & 16.97 & 4.8642 & 4.9581 & 7.8971 & 6.8332\\
      & 24 & 18.25 & 17.85 & 4.9576 & 4.9001 & 7.0551 & 6.0549   \\
\hline
  \raisebox{-6pt}{1.1} & 20 & 15.05 & 15.45 & 5.5735 & 5.7477 & 6.2795 & 7.9492\\
      & 24 & 16.50 & 16.19 & 5.7496 & 5.6340 & 8.3960 & 6.7278  \\

\hline
  \raisebox{-6pt}{1.2} & 20 & 15.05 & 14.23 & 5.5734 & 7.2771 & 6.2792 & 9.0045\\
      & 26 & 15.30 & 14.90 & 6.9276 & 6.4566 & 8.9663 & 8.7410  \\[2pt]
\hline
\end{tabular}
\end{center}
\end{table}

\begin{table}
\caption{Average gap characteristics in the P3 model}
\bigskip
\begin{tabular}{|c|c|c|c|c|c|c|c|c|c|c|c|c|}
\hline\rule{0pt}{14pt}
  {$k_{\rm F},\!\!$} & \raisebox{-6pt}{$Z$} &
  \multicolumn{2}{|c|}{$k_{\rm F}^{\rm as},\,{\rm fm^{-1}}$}&
  \multicolumn{2}{|c|}{$\Delta(0),\;$MeV}&
  \multicolumn{2}{|c|}{$\Delta_{\rm as},\;$MeV}&
  \multicolumn{2}{|c|}{$\Delta_{\rm F},\;$MeV}&
  \multicolumn{2}{|c|}{$\Delta_{\rm inf},\;$MeV}&
  {$\Delta^0_{\rm inf},\!\!$}\\
\cline{3-12}
 \rule{0pt}{13pt}$\rm fm^{-1}\!\!$&& BC1 & BC2 & BC1 & BC2 & BC1 & BC2 & BC1 & BC2 & BC1 &  BC2
 &MeV$\!\!$\\
 \hline\rule{0pt}{13pt} \raisebox{-6pt}{0.6}& 56 &0.5817 &0.5788 &0.721 & 0.755 & 0.719 & 0.713 &0.723
&0.715 & 0.778 &0.774 &  \raisebox{-6pt}{0.81}\\
 & 54 & 0.5787 &0.5792 &0.755 &0.701 &0.732 &0.704
&0.722 &0.713 &0.774 &0.775 &\\

\hline \raisebox{-6pt}{0.7} & 46 &0.6752 &0.6782 &0.789 &0.850
&0.832 &0.852
&0.806 &0.865 &0.893 &0.896 &\raisebox{-6pt}{0.92} \\
& 48 & 0.6826 & 0.6739 &0.924 &0.765 &0.864 &0.797 &0.864 &0.817 &0.901 &0.891 &  \\
\hline\rule{0pt}{13pt} {0.8} &40 &0.7676 & 0.7791 & 0.699 & 1.019
&0.787 & 0.923 &0.799 &0.943  &0.958 &0.964 &0.91\\
\hline \raisebox{-6pt}{0.9} &20 & 0.8489 &0.8828 &0.080 &1.019
&0.107 &0.887 &0.097 &0.975
&0.974 &0.972 &\raisebox{-6pt}{0.90}\\
& 24 &0.8747 &0.8450 &0.995 &0.367 &0.817 &0.468 &0.919 &0.479 &0.973 &0.974 & \\
\hline \raisebox{-6pt}{1.0} &20 &0.9399 &0.9690 &0.011 &1.335
&0.019 &0.919 &0.017 & 1.100
&0.941 &0.919 & \raisebox{-6pt}{0.89} \\
& 24 &0.9636 &0.9323 &1.291 &0.134 &0.909 &0.167 &1.068 &0.174 &0.923 &0.947 &\\
\hline \raisebox{-6pt}{1.1} &20 &1.0315 &1.060 &0.035 &1.472
&0.028 &0.907
&0.028 & 1.154 &0.850 &0.810 &\raisebox{-6pt}{0.75}\\
& 24 &1.0583 &1.0261 &1.236 &0.618 &0.899 &0.375 &1.10 &0.419 &0.812 &0.858 & \\
\hline \raisebox{-6pt}{1.2} &20 &1.1254 &1.1458 &1.154 &0.010
 &0.313 &0.013
&0.418 &0.018 &0.702 &0.661 &\raisebox{-6pt}{0.55}\\
& 26 &1.1308 &1.1146 &0.157 &0.111 &0.239 & 0.021 &0.270 &0.031
&0.691 &0.724&\\[2pt]
 \hline
\end{tabular}
\end{table}
\newpage

\begin{figure}
\centerline{\includegraphics [height=100mm,width=120mm]{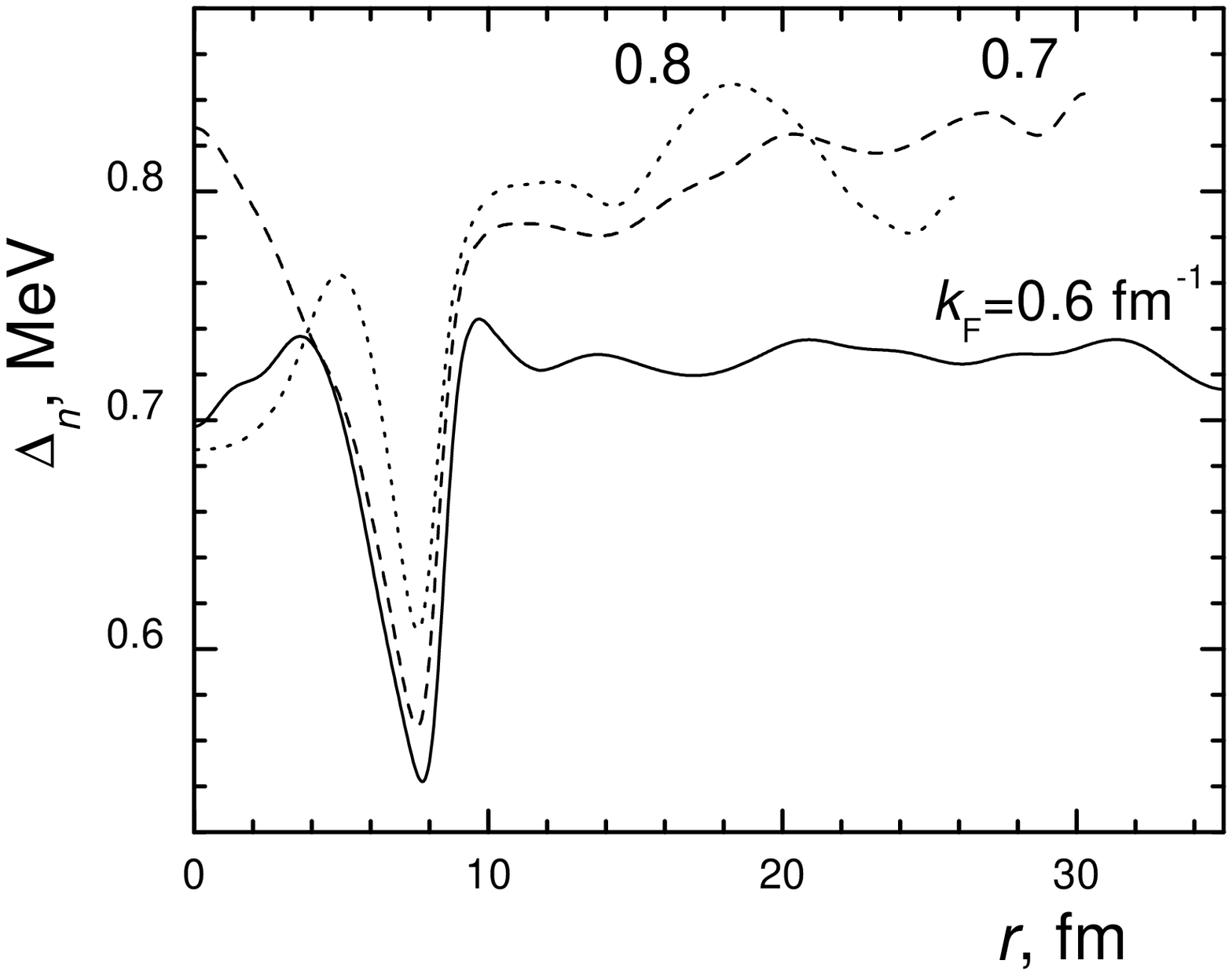}}
\vspace{2mm} \caption{The neutron gap $\Delta_n(r)$
 for $k_{\Fs}{=}0.6 \div 0.8  \;$fm$^{-1}$ in the case
of the P3 model.}
\end{figure}

\begin{figure}
\centerline{\includegraphics [height=100mm,width=120mm]{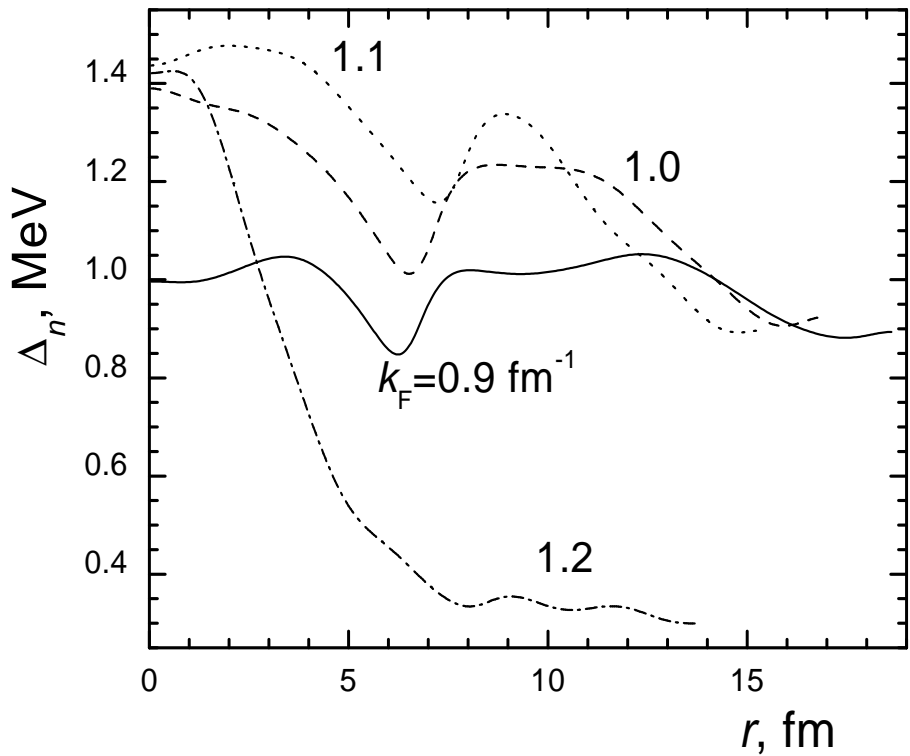}}
\vspace{2mm} \caption{The neutron gap $\Delta_n(r)$
 for $k_{\Fs}{=}0.9 \div 1.2  \;$fm$^{-1}$ in the case
of the P3 model }
\end{figure}

\section{ Discussion and Conclusions}

  Recently a semi-microscopic self-consistent quantum  approach was developed
\cite{crust3,crust4}  for describing the inner crust structure of
neutron stars within the WS method with taking into account
pairing correlation effects. It is based on the generalized energy
functional method \cite{Fay} which is a modified version of the
original Kohn-Sham one \cite{KS} for the case of superfluid
systems. In this approach,  the energy functional is constructed
by matching the realistic phenomenological functional by Fayans et
al. \cite{Fay} for describing the nuclear-type cluster in the
center of the WS cell to the one calculated microscopically for
neutron matter. The anomalous  part of the latter was calculated
within the BCS approximation. In this paper we take into account,
in an approximate way, corrections to the BCS theory which are
known from the many-body theory of pairing in neutron matter.

Unfortunately, up to now there is no consistent many-body theory
of pairing in neutron matter. However, there exists a conventional
point of view  \cite{crust2} that the BCS gap value is suppressed
due to various many-body theory corrections significantly, by a
factor between 1/2 and 1/3. In the method developed in
\cite{crust3,crust4}, the set of the neutron matter gap values
$\Delta_n(k_{\Fs})$ at the Fermi surface for the interval of $0<
k_{\Fs}<1.35\;$fm$^{-1}$ is the only input to the microscopic part
of the superfluid component of the energy functional. In fact, we
limit ourselves with the interval of 0.6 fm$^{-1} < k_{\Fs}<1.2
\;$fm$^{-1}$ in which the neutron pairing effects are expected to
be larger. It is worth to note that the values of $ k_{\Fs} \ge
1.1 \;$fm$^{-1}$ should be considered as optional as far as the WS
configuration with spherical symmetry is evidently unstable in
this density region \cite{Oyam,Mag}.

We use a simple model to take into account approximately the
many-body corrections to the BCS theory. In this model, the BCS
value $\Delta_n^{\rm BCS}(k_{\Fs})$ is suppressed by a density
independent factor which was taken to be $f_{\rm mb}{=}1/2$ in the
first version of the model (named P2) and $f_{\rm mb}{=}1/3$  in the
second one, P3. These corrections influence the equilibrium
configurations ($Z,R_c$) at different $k_{\Fs}$. The maximal
variation from the BCS theory (the P1 model) to the P2 version
occurs at $ k_{\Fs}{=}0.7 \;$fm$^{-1}$, the equilibrium Z value
changing by 6 units.  As to the difference of the ($Z,R_c$)
configurations found within the P2 and P3 models for the same
version of the boundary conditions, it is usually negligible. The
most important variation for P2 {\it versus} P3 occurs in the
neutron gap function itself. We think that the realistic situation
takes place somewhere between the P2 and P3 models.

\section{ Acknowlegments }

This research was partially supported by the Grant NSh-8756.2006.2
of the Russian Ministry for Science and Education and by the RFBR
grant 06-02-17171-a.

{}

\end{document}